\documentclass[prl,
,tightenlines
,showkeys
,showpacs
]
{revtex4-2}

\usepackage{amsmath,amssymb}
\usepackage{bm}
\usepackage[dvipdfmx]{graphicx}
\usepackage{ascmac} 
\usepackage{fancybox}
\usepackage{float}
\usepackage{enumerate}
\usepackage{color}
\usepackage{amsthm}
\usepackage{mathrsfs}
\usepackage{comment}
\usepackage{ulem}
\allowdisplaybreaks[1]
\usepackage{comment}
\newtheorem*{th.}{Theorem}

\begin{document}
\title{Heisenberg-limited quantum metrology using collective dephasing}
\date{\today}
\author{Shingo Kukita$^{1)}$}
\email{toranojoh@shu.edu.cn}
\author{Yuichiro Matsuzaki$^{2)}$}
\email{matsuzaki.yuichiro@aist.go.jp}
\author{Yasushi Kondo$^{1)}$}
\email{ykondo@kindai.ac.jp}
\affiliation{$^{1)}$Department of Physics, 
Kindai University, Higashi-Osaka 577-8502, Japan}
\affiliation{$^{2)}$Device Technology Research Institute,
National Institute of Advanced Industrial Science and Technology (AIST),
1-1-1, Umezono, Tsukuba, Ibaraki 305-8568, Japan}
\begin{abstract}

The goal of quantum metrology 
is the precise estimation of
parameters 
using quantum properties such as entanglement. This estimation usually consists of three steps: state preparation, time evolution during which information of the parameters is encoded in the state, 
and
readout of the state. Decoherence during the time evolution 
typically degrades the performance of quantum metrology and is considered to be 
one of the major obstacles to realizing entanglement-enhanced sensing.
We show, however, that under suitable conditions,
this decoherence
can be exploited to improve the sensitivity.
Assume that we have two axes,
and our aim is 
to estimate the relative angle between them.
Our results reveal that 
the use of Markvoian
collective dephasing to estimate 
the relative angle between the two directions
affords Heisenberg-limited sensitivity. Moreover, our scheme based on Markvoian collective dephasing is robust against environmental noise, and
it is possible to achieve the Heisenberg limit even under the effect of independent dephasing.
Our counterintuitive results showing that the sensitivity is improved by using the decoherence
pave the way to novel applications in quantum metrology.

\end{abstract}
\maketitle

Sensing technology is important for many practical applications \cite{huber2008gradiometric,ramsden2011hall,poggio2010force},
and improved sensitivity is essential for practical purposes.
Quantum metrology
is a promising approach in order to improve the sensitivity 
using qubits
owing to recent developments in quantum technology \cite{helstrom1976quantum,dunningham2006using,holevo2011probabilistic,caves1981quantum,giovannetti2004quantum,giovannetti2006quantum,simon2017quantum,giovannetti2011advances,taylor2016quantum,degen2017quantum,paris2009quantum}.
Quantum states can acquire a phase during interaction with the target fields.
The readout of the phase provides information on the amplitude of the target fields \cite{wineland1992spin,wineland1994squeezed,toth2014quantum,bollinger1996optimal,macieszczak2015zeno,PhysRevLett.79.3865,gorecka2018noisy}.
Quantum sensing allows us to measure not only the amplitude of the fields but also many other quantities.
Parameters that can be measured using qubit-based sensing include the Fourier coefficients of the spatially distributed fields \cite{rossi2020noisy}, field gradient \cite{altenburg2017estimation}, frequency of AC magnetic fields \cite{schmitt2017submillihertz}, and rotation \cite{ledbetter2012gyroscopes,ajoy2012stable}.
When $n$ separable qubits are used as a probe,
the uncertainty of parameter estimation scales as ${\cal O}(1/\sqrt{n})$,
which is called the standard quantum limit (SQL).
By contrast, the uncertainty scales as
$\sim{\cal O}(1/n)$ when highly entangled
states of qubits, 
such as Greenberger-Horne-Zeilinger (GHZ) states, are used
\cite{monz201114,greenberger1990bell,dicarlo2010preparation}.
This scaling 
is called the Heisenberg limit (HL) \cite{holland1993interferometric,bollinger1996optimal,giovannetti2006quantum}.
Many
studies have been conducted to achieve Heisenberg-limited sensitivity
\cite{nagata2007beating,jones2009magnetic,facon2016sensitive,kruse2016improvement,cox2016deterministic,hosten2016measurement,luo2017deterministic,mason2019continuous,bao2020spin,pedrozo2020entanglement}.

In realistic situations, 
entangled qubits are affected by 
environmental noise during the time evolution required to encode the parameter information, and this decoherence is one of the main obstacles to realizing entanglement-enhanced sensors.
If the noise acts independently on the qubits, the entanglement of the qubits rapidly disappears,
and the states of the $n$ qubits become separable.
Thus, it is not trivial whether entanglement sensors are useful.
Numerous attempts have been made to address the problem of decoherence in order to overcome the SQL with entangled sensors
\cite{PhysRevLett.79.3865,PhysRevA.76.032111,matsuzaki2011magnetic,chin2012quantum,demkowicz2012elusive,chaves2013noisy,tanaka2015proposed,davis2016approaching,matsuzaki2018quantum1}.
Measurements in a quantum Zeno regime can be adopted to achieve a scaling of ${\cal O}(n^{3/4})$ if the noise is time-inhomogeneous independent dephasing
\cite{matsuzaki2011magnetic,chin2012quantum,matsuzaki2018quantum1,macieszczak2015zeno,gorecka2018noisy,hakoshima2020multiparameter,yoshinaga2021entanglement}.
In addition, quantum error correction can be applied to noisy metrology to suppress the effect of decoherence
\cite{kessler2014quantum,arrad2014increasing,dur2014improved,herrera2015quantum,matsuzaki2017magnetic},
and this method has been demonstrated by several experiments \cite{unden2016quantum,cohen2016demonstration}.
Quantum teleportation is another tool that protects 
quantum states from
the effects of noise \cite{matsuzaki2018quantum1,matsuzaki2016hybrid,averin2016suppression}.
There is a scheme for reaching the HL in the estimation of the decay rate using dephasing \cite{beau2017nonlinear,matsuzaki2018quantum2}.
Measurements of the environment itself improve the sensitivity of parameter estimation even under the effect of noise \cite{braun2011heisenberg}.
There are several other methods for improving the sensitivity of estimation under noise
\cite{demkowicz2017adaptive,sekatski2017quantum,dooley2018robust,koczor2020variational,rossi2020noisy}.

\begin{figure}
\begin{center}
\includegraphics[width=85mm]{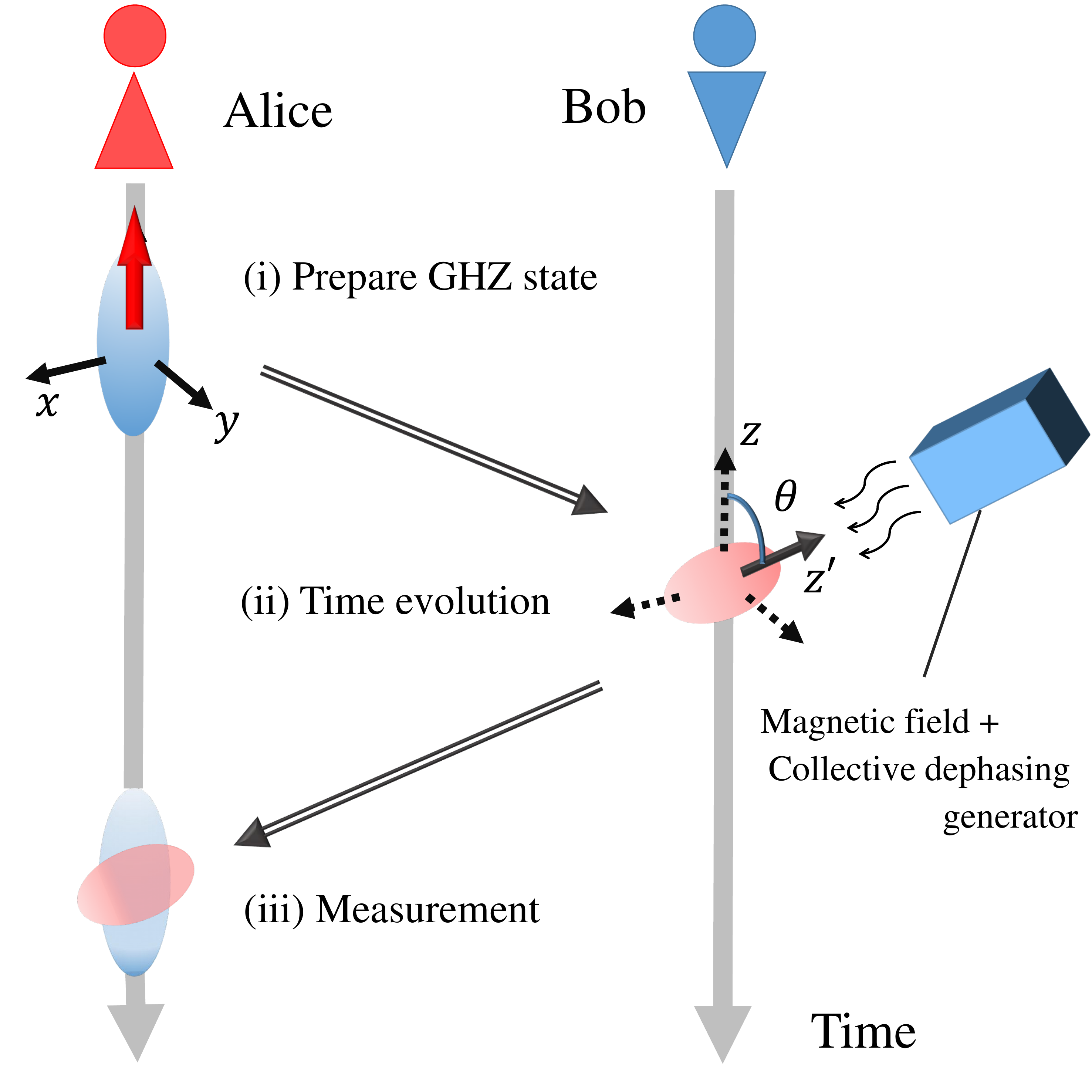}
\caption{Schematic illustration of the proposed protocol. (i) Alice prepares a GHZ state, (ii) Bob receives this state and lets it evolve under the applied collective noise (or a global magnetic field), and (iii) Alice measures this state. 
}
\label{fig:AliceBob}
\end{center}
\end{figure}

In this paper, we propose a quantum metrology protocol using collective dephasing 
to enhance
the sensitivity. 
Suppose that Alice has an axis 
and Bob has another axis.
Bob does not know Alice's axis
and tries to estimate the relative angle between her axis and his own.
The setup is as follows (Fig.~\ref{fig:AliceBob}).
(i) Alice prepares qubits in a GHZ state according to her
axis
and sends the qubits to Bob.
(ii) Bob applies global magnetic fields or the collective dephasing noise
along his
axis on the qubits he received
and sends them back
to Alice.
(iii) Alice reads out the state and 
sends the measurement results to Bob by classical communication.
(iv) They repeat these three steps $M$ times.
We have $M=T/(t_{\rm{prep}}+t_{\rm{evolve}}+t_{\rm{read}})$, where $T$ denotes the total time allowed for the protocol, $t_{\rm{prep}}$ denotes the time needed to prepare the GHZ state (which includes the transportation time), $t_{\rm{evolve}}$ denotes the evolution time, and $t_{\rm{read}}$ denotes the time required to read out the state. Throughout this paper, we assume that the GHZ state can be prepared and read out on a much shorter time scale than the evolution time, and we obtain $M \simeq T/t_{\rm{evolve}}$.

Let us describe the details of our setup. We define Alice's (Bob's)
axis as the $z$ ($z'$) axis.
In Step (i), Alice prepares $n$ qubits in a GHZ state, 
which is defined as follows.
\begin{equation}
|{\rm GHZ}\rangle=\frac{1}{\sqrt{2}}(|\underbrace{\uparrow\uparrow\cdots\uparrow}_{n}\rangle+|\underbrace{\downarrow\downarrow\cdots\downarrow}_{n}\rangle),
\end{equation}
where $|\uparrow \rangle$ ($|\downarrow \rangle$) is the eigenstate of $\sigma_{z}$ with an eigenvalue of $+1~(-1)$,
and $|\uparrow\uparrow\cdots\uparrow\rangle$ denotes $|\uparrow\rangle \otimes |\uparrow\rangle \otimes\cdots \otimes |\uparrow\rangle$.
Here we take the ordinary notation of the Pauli matrices.
Note that the $x$ and $y$ axes are actually fixed when the relative phase in the GHZ state is fixed.

In Step (ii), to encode the information on the relative angle, Bob can apply a global magnetic field or the collective dephasing noise along the $z'$ axis to the GHZ state 
that he receives from Alice.
In addition, we assume that environmental Markovian dephasing noise independently affects each qubit along the $z'$ axis.
We introduce the vector $\vec{z'}=(\sin \theta\cos \phi,\sin \theta\sin \phi,\cos \theta)$, which is the unit vector along the $z'$ direction represented in the ($x$,$y$,$z$) coordinates of Alice.
$\theta$ is the parameter to be estimated.
The Pauli matrix along the $z'$ direction is written as $\sigma_{z'}=\vec{z'}\cdot\vec{\sigma}$.
In addition, we use the notation $\sigma^{(i)}_{\alpha}$ ($\alpha=x,y,z$)
for
a Pauli matrix acting only on the $i$-th qubit, e.g., $\sigma^{(1)}_{\alpha}=\sigma_{\alpha}\otimes {\mathbb I}\cdots \otimes {\mathbb I}$,
where ${\mathbb I}$ is the $2\times 2$ identity matrix.
Thus, the dynamics of the GHZ state
on Bob's side is described as follows:
\begin{align}
\frac{d \rho}{d t}=& -i [\Omega L_{z'},\rho]+\gamma \Bigl( L_{z'} \rho L_{z'}-\frac{1}{2}\{L_{z'}^{2} ,\rho\}\Bigr)\nonumber\\
&+\gamma' \sum^{n}_{i=1}(\sigma_{z'}^{(i)}\rho\sigma_{z'}^{(i)}-\rho),
\label{eq:lindblad}
\end{align}
where $L_{z'}=\sum^{n}_{i=1}\sigma_{z'}^{(i)}$,
and $\Omega$ characterizes the strength of the global magnetic field.
Throughout this paper, we take $\hbar=1$.
Bob can tune the values of $\gamma$ and $\Omega$, whereas $\gamma '$ is not tunable.

Our goal is to estimate the azimuthal angle $\theta$ with high precision.
We take $\phi=0$ for simplicity.
Note that the exact solution of Eq.~(\ref{eq:lindblad}) is analytically given, and we show that our protocol for estimating $\theta$ does not depend on the value of $\phi$ in the parameter regime of interest. (See Supplemental Material.)
We focus on the case of $(\Omega=0,\gamma\neq0,\gamma'=0)$, $(\Omega=0,\gamma\neq0,\gamma'\neq0)$, $(\Omega\neq0,\gamma=0,\gamma'=0)$, and $(\Omega\neq0,\gamma=0,\gamma'\neq0)$ to evaluate the advantages of our scheme
using
collective dephasing over that using the global magnetic field.

For $(\Omega=0,\gamma\neq0,\gamma'=0)$ and $(\Omega=0,\gamma\neq0,\gamma'\neq0)$,
where Bob uses collective dephasing,
Alice performs a projective measurement defined by the operator $\hat{\mathcal{P}}=|{\rm GHZ}\rangle\langle{\rm GHZ}|$. The projection using this operator provides a
survival probability $P:=\langle {\rm GHZ}|\rho(t)|{\rm GHZ}\rangle$ in step (iii).
Then, Bob estimates the value of $\theta$ by analyzing the $M$ outcomes of the projections.
The uncertainty of this estimation, $\delta \theta$, is bounded by
\begin{equation}
\delta \theta\geq\delta \theta^{\rm min}=\frac{\sqrt{P(1-P)}}{|\frac{d P}{d \theta}|\sqrt{M}}= \frac{\sqrt{P(1-P)}}{|\frac{d P}{d \theta}|\sqrt{\frac{T}{t_{\rm evolve}}}}.
\label{eq:bound}
\end{equation}
The lower bound $\delta \theta^{\rm min}$ depends on the evolution time $t_{\rm evolve}$.
Hence, we need to optimize 
$t_{\rm evolve}$ so that $\delta \theta^{\rm min}$ 
takes
the smallest value. 
We find below that $|{\rm GHZ}\rangle \langle{\rm GHZ}|$ is an appropriate measurement operator in this case; i.e., the minimized uncertainty as defined above achieves the HL.

By contrast, in the
scheme of applying the global magnetic field, $(\Omega\neq0,\gamma=0,\gamma'=0)$ and $(\Omega\neq0,\gamma=0,\gamma'\neq0)$,
the projection operator of $|{\rm GHZ}\rangle \langle{\rm GHZ}|$ is not the optimal choice.
For mixed states, it is not trivial to find the optimal positive-operator-valued measure (POVM) to minimize the uncertainty.
Hence, we employ the minimized uncertainty $\delta \theta^{\rm (Q) min}=1/(F^{(Q)}_{\theta}\sqrt{T/t_{\rm evolve}})$ defined by the quantum Fisher information $F_{\theta}^{(Q)}$. (See Supplemental Material for the definition.)
This minimum $\delta \theta^{\rm (Q) min}$ corresponds to the minimized uncertainty
when we adopt the best POVM.
Importantly,
we can calculate this minimum without knowing the best measurement basis.

\begin{figure*}
\centering
\hspace{-0.5cm}
\includegraphics[width=89mm]{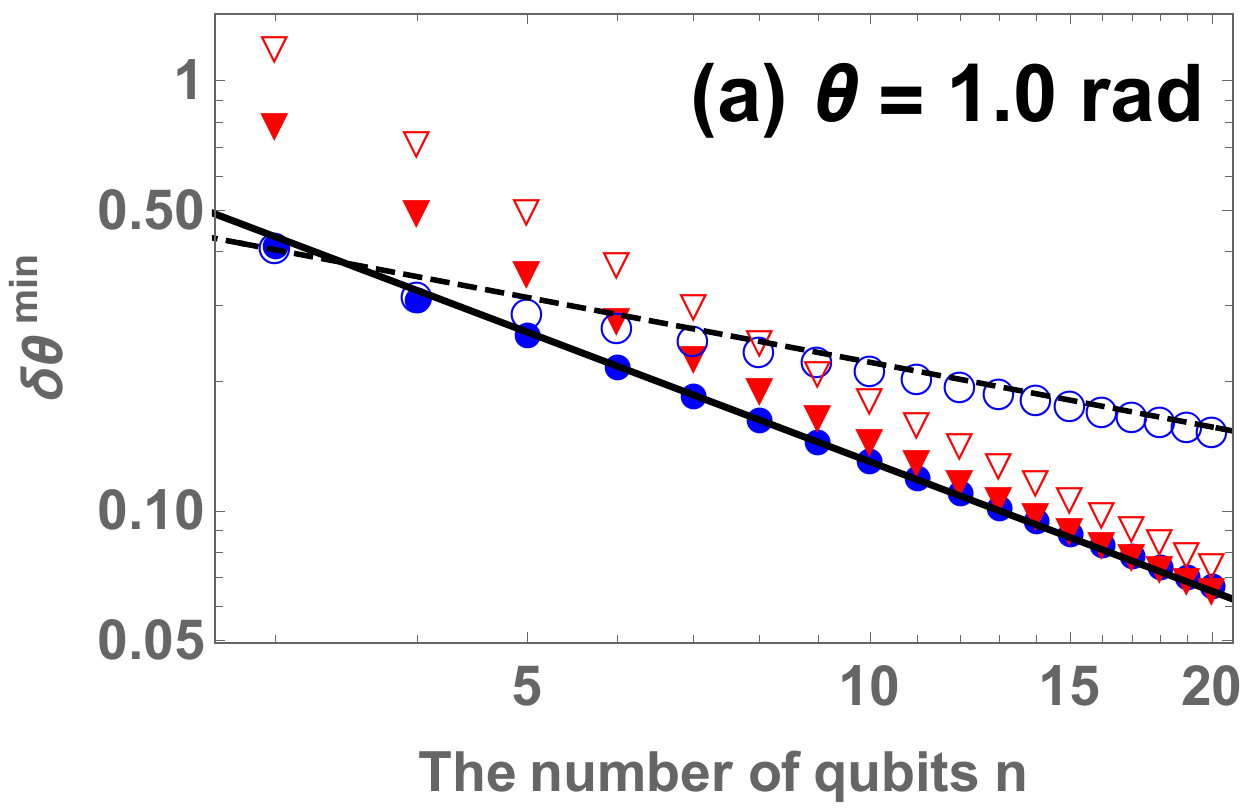}
\includegraphics[width=87mm]{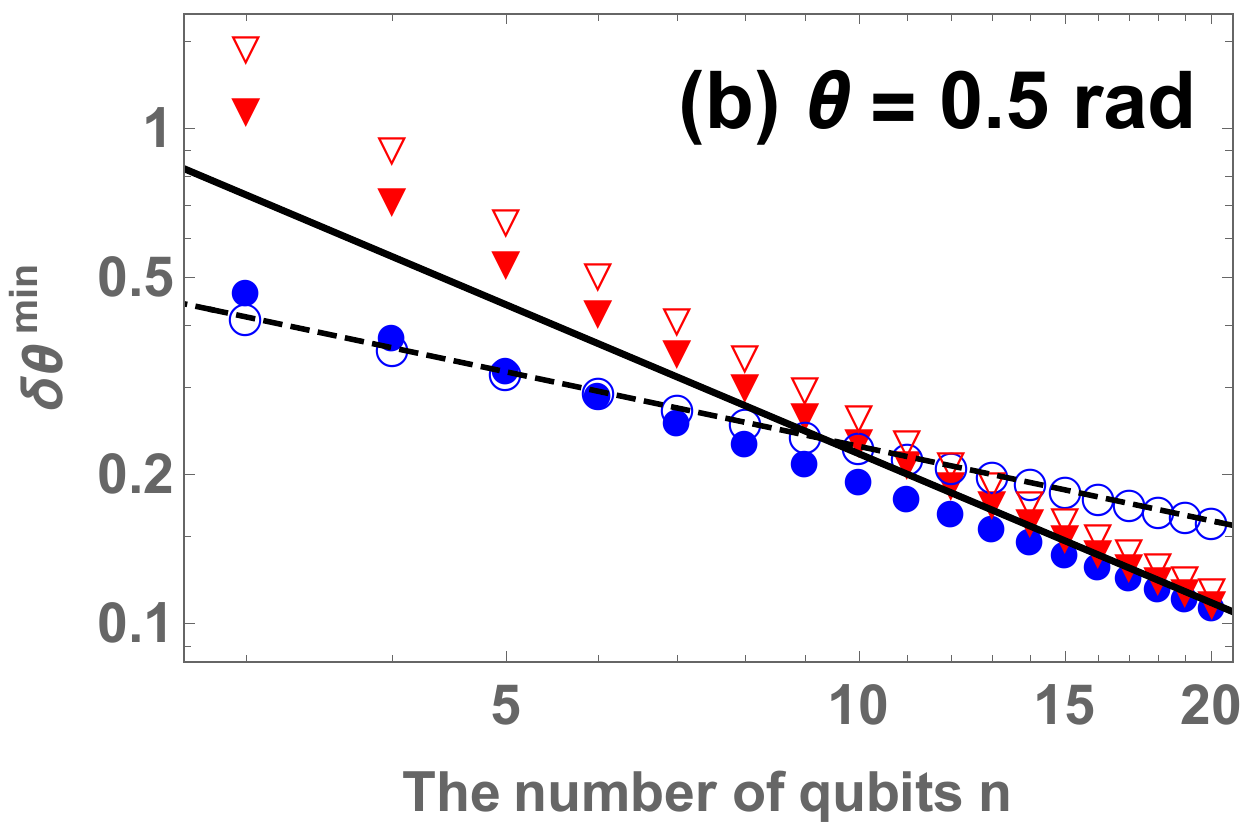}
\caption{
$\delta \theta^{\rm min}_{1}$ ($\delta \theta^{\rm (Q) min}_{1}$) versus the number of qubits $n$
for (a) $\theta=1.0$ rad and (b) $\theta=0.5$ rad.
In both panels, the filled (open) triangles represent
$\delta \theta^{\rm min}_{1}$
with the parameters $\Omega=0$, $\gamma=1$, and $\gamma'=0~(1)$,
whereas the filled (open) circles 
represent
$\delta \theta^{\rm (Q) min}_{1}$
with the parameters $\Omega=1$, $\gamma=0$, and $\gamma'=0~(1)$.
The solid (dashed) line shows the HL (SQL). 
The total time $T$ is taken as $T=1$.
}
\label{fig:result1}
\end{figure*}

\begin{figure*}
\begin{center}
\hspace{-1.4cm}
\includegraphics[width=90mm]{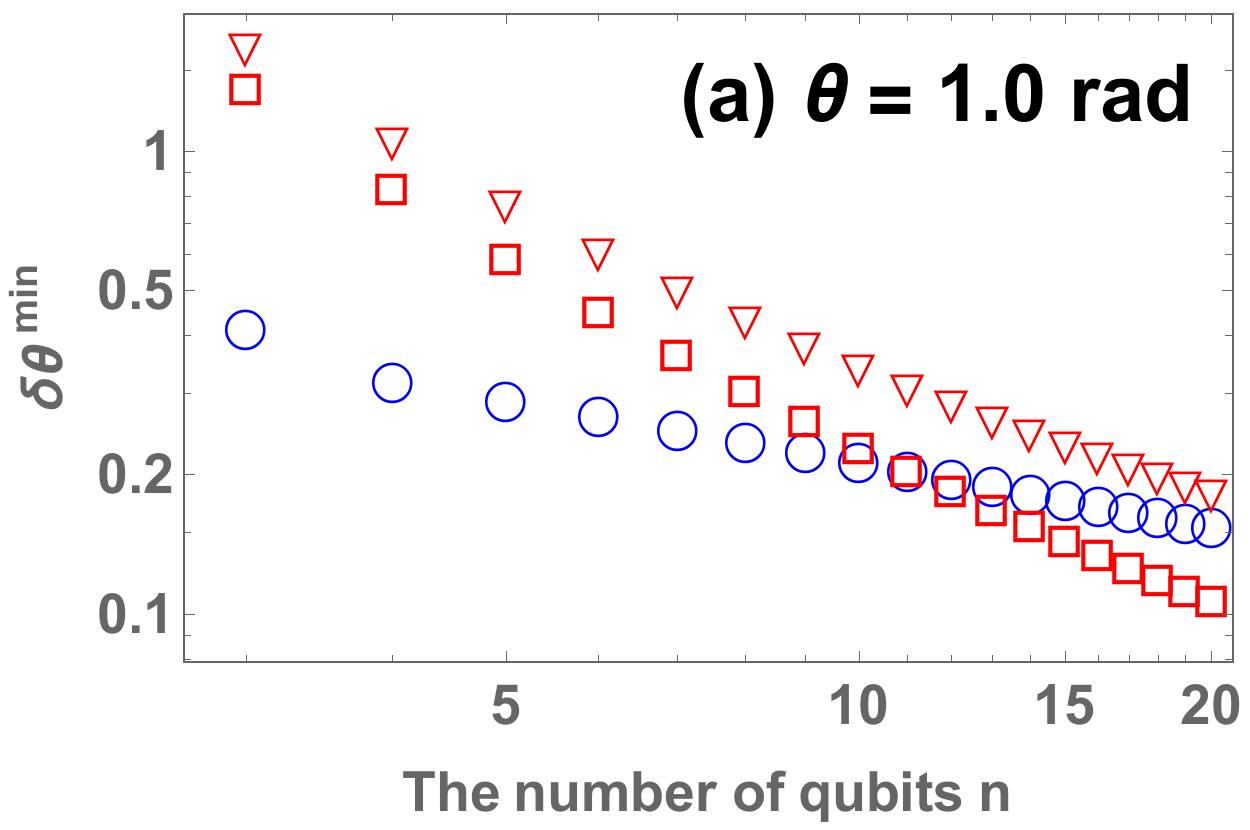}
\includegraphics[width=90mm]{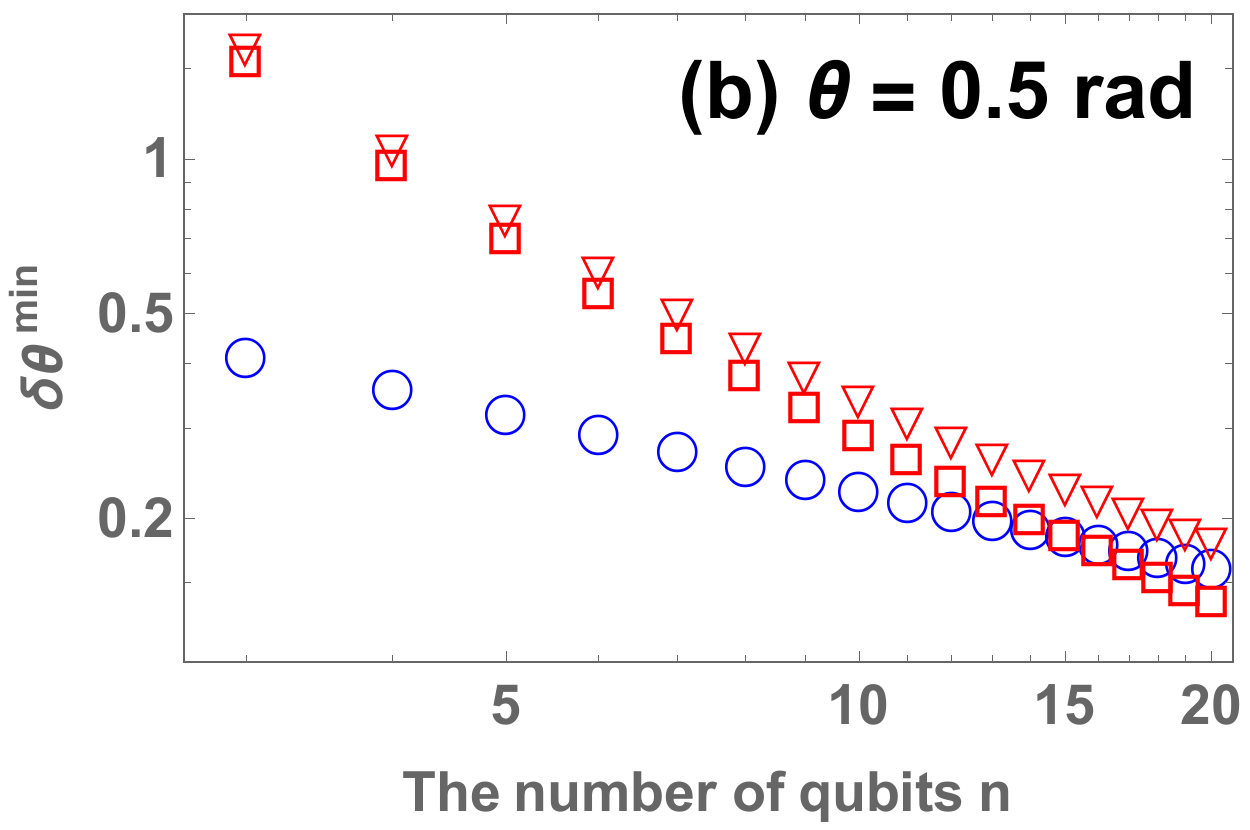}
\hspace{-1.2cm}
\caption{
Minimized uncertainty
$\delta \theta^{\rm min}$ ($\delta \theta^{\rm (Q) min}$) versus the 
number of qubits $n$ for non-Markovian collective dephasing
for (a) $\theta=1.0$ rad and (b) $\theta=0.5$ rad.
In both panels, the blue circles represent
$\delta\theta^{\rm (Q) min}$ with parameters $\Omega=1$, $\gamma_{0}=0$, and $\gamma'=1$, which give the same results as in Fig. \ref{fig:result1} (a) and (b).
In (a), the red triangles (squares) represent
the uncertainty with parameters $\Omega=0$, $\gamma_{0}=1$, $\tau_{c}=0.01 (0.001)$, and $\gamma'=1$,
whereas the red triangles (squares) in (b) show
the uncertainty with parameters $\Omega=0$, $\gamma_{0}=1$, $\tau_{c}=0.005~(0.0001)$, and $\gamma'=1$.
}
\label{fig:result2}
\end{center}
\end{figure*}

Figure \ref{fig:result1}
shows the scaling behavior of the minimized uncertainty
versus the number of qubits
$n$ for $(\Omega,\gamma,\gamma')=(0,1,0)$, $(0,1,1)$, $(1,0,0)$, and $(1,0,1)$.
Figure \ref{fig:result1} (a) and (b) correspond to the case where we take $\theta=1.0$ rad and $\theta=0.5$ rad, respectively.
In the noiseless cases, $(\Omega,\gamma,\gamma')=(1,0,0)$ and $(0,1,0)$
in Fig. \ref{fig:result1}, we 
find
that the minimized uncertainties
in both cases 
approach 
the HL for large $n$.
However, estimation using
the magnetic field
is fragile against independent dephasing,
as shown
in Fig. \ref{fig:result1},
where
the optimized uncertainty scales as the SQL.
By contrast, 
estimation using collective dephasing is robust against independent dephasing, 
as shown in Fig. \ref{fig:result1},
and thus the estimation scheme using collective dephasing outperforms that using the global magnetic field in this case.
Note that a specific measurement basis
($|{\rm GHZ}\rangle \langle{\rm GHZ}|$)
is chosen
for estimation using collective dephasing;
the uncertainty of estimation using the global magnetic field is evaluated on the basis of the quantum Fisher information without knowledge of the explicit form of the POVM to employ.
If we could find the optimized measurement basis for estimation using collective dephasing, we could improve the sensitivity by a constant factor.
Moreover, by using perturbative calculations, we show analytically that
the minimized uncertainties of collective dephasing
approaches the HL
even under the effect of independent dephasing.
Here, the optimal evolution
time scales as $\propto 1/n^{2}$. (See Supplemental Material.)

In quantum metrology, the sensitivity under Markovian noise could be very different from that under non-Markovian (time-inhomogeneous) noise \cite{PhysRevLett.79.3865,matsuzaki2011magnetic,chin2012quantum}.
The non-Markovian noise model takes into account the finite correlation time of the environment, whereas the Markovian environment has an infinitesimal correlation time. Owing to the finite correlation time, a typical non-Markovian noise model interpolates between exponential decay (which is typically observed in Markovian noise) and quadratic decay.

We investigate the sensitivity of our scheme when we use 
non-Markovian collective dephasing for estimation.
In particular,
we adopt a spin-boson model with a Lorentzian spectral density to consider the effect of the finite correlation time.
This model was analyzed in \cite{chin2012quantum}, and the time-dependent decay rate was calculated as $\gamma(t)=\frac{\gamma_{0}\tau_{c}}{t}(-1+e^{-t/\tau_{c}}+t/\tau_{c})$, where $\tau_{c}$ denotes the correlation time.
This decay rate interpolates between exponential decay
and quadratic decay.
For a short (long)
correlation time, we obtain $\gamma(t) \simeq \gamma_{0}$ ($\gamma(t) \simeq\frac{\gamma_{0} t}{2\tau_{c}}$).

We compare the uncertainty of the estimation using the non-Markovian collective dephasing with that using global magnetic fields
by performing numerical simulations.
The results are shown in Fig. \ref{fig:result2} (a) and (b), where we take $\theta=1.0$ rad and $\theta=0.5$ rad, respectively.
Figure \ref{fig:result2} shows that the estimation using collective dephasing outperforms that using global magnetic fields when we take a sufficiently small $\tau_{c}$.
In the numerical simulations, we observe that either the collective dephasing method or the global magnetic field method approaches the SQL. (This behavior is also discussed using an analytical calculation in the Supplemental Material.)
Whether the use of the non-Markovian collective dephasing is advantageous over the use of the global magnetic field depends on both $\tau_{c}$ and $\theta$.
For $\theta=1.0$ rad, $\tau_{c}=0.001$ is sufficiently small, whereas $\tau_{c}\sim 0.0001$ is required for $\theta=0.5$ rad.
We emphasize that the estimation scheme
using collective dephasing can outperform that using the magnetic field for any $\theta$ if we take sufficiently small $\tau_{c}$, because the scaling behavior achieves the HL in the Markovian limit $\tau_{c}\rightarrow 0$.

In conclusion, we propose to use a collective dephasing to improve the precision of quantum metrology.
Assume that we have two axes, 
and our aim is 
to estimate the relative angle between them.
Suppose that Alice has an axis, 
and Bob has another.
Bob does not known Alice's 
and tries to estimate the relative angle between her axis and his.
Alice generates a GHZ state according to her
axis and sends it to Bob.
Bob decoheres the received state 
by inducing collective Markovian dephasing along his own 
axis.
This scheme
achieves the HL for estimating the direction of Alice's
axis
under ideal conditions.
Moreover, we show that the scheme using collective dephasing is robust against noise; it
achieves the HL even under the effect of independent Markovian dephasing from the environment.
This is in stark contrast to the conventional scheme that uses unitary dynamics for the estimation, which cannot overcome the SQL under the effect of such noise.
Although we discuss primarily the independent dephasing noise, our conclusion that the HL can be achieved is guaranteed even when the system is affected by arbitrary types of independent decoherence. (See Supplemental Material).

This work was supported by the Leading Initiative for
Excellent Young Researchers, MEXT, Japan; JST
Presto (Grant No. JPMJPR1919) Japan; and CREST (JPMJCR1774).

\section{Supplemental Material}
\subsection{Exact solution of the Lindblad equation}

Let us first introduce basic notation.
We refer to Alice's coordinates as the $(x,y,z)$ coordinates, and the corresponding Pauli matrices are given as
\begin{equation}
\sigma_{x}=
\begin{pmatrix}
0&1\\
1&0
\end{pmatrix}
,~
\sigma_{y}=
\begin{pmatrix}
0&-i\\
i&0
\end{pmatrix}
,~
\sigma_{z}=
\begin{pmatrix}
1&0\\
0&-1
\end{pmatrix}
.
\end{equation}
The explicit form of $\sigma_{z'}=\vec{z'}\cdot\vec{\sigma}$
in Bob's coordinates is as follows:
\begin{equation}
\sigma_{z'}=\vec{z'}\cdot \vec{\sigma}=
\begin{pmatrix}
\cos \theta&e^{-i \phi}\sin \theta\\
e^{i \phi}\sin \theta&-\cos \theta
\end{pmatrix}
.
\end{equation}
The corresponding eigenstates are defined as
\begin{align}
|\uparrow\rangle_{z'}=&\frac{1}{\sqrt{2}}\Bigl(\sqrt{1+\cos \theta}~|\uparrow\rangle+e^{i \phi}\sqrt{1-\cos \theta}~|\downarrow\rangle \Bigr),~~\sigma_{z'}|\uparrow\rangle_{z'}=|\uparrow\rangle_{z'},\nonumber\\
|\downarrow\rangle_{z'}=&\frac{1}{\sqrt{2}}\Bigl(-\sqrt{1-\cos \theta}~|\uparrow\rangle+e^{i \phi}\sqrt{1+\cos \theta}~|\downarrow\rangle \Bigr),~~\sigma_{z'}|\downarrow\rangle_{z'}=-|\downarrow\rangle_{z'}.
\end{align}
Here $|\uparrow\rangle$ and $| \downarrow \rangle$ are the eigenvectors of $\sigma_{z}$, whose eigenvalues are $1$ and $-1$, respectively.

Here we give the exact solution of the Lindblad master equation,
\begin{equation}
\frac{d \rho}{d t}= -i [\Omega L_{z'},\rho]+\gamma \Bigl( L_{z'} \rho L_{z'}-\frac{1}{2}\{L_{z'}^{2} ,\rho\}\Bigr)+\gamma' \sum^{n}_{i=1}(\sigma_{z'}^{(i)}\rho\sigma_{z'}^{(i)}-\rho).
\label{eq:lindbladsup}
\end{equation}
Let us consider the case of $\gamma'=0$ first.
To this end, it is convenient to introduce a basis according to group representation theory \cite{mihailov1977addition,ping2002group,chase2008collective,baragiola2010collective}, which is characterized as follows:
\begin{align}
|j,m,i\rangle_{z(z')}&\in {\mathbb C}^{2 n},~j_{\rm min}\leq j \leq n/2,~-j\leq m \leq j,~1 \leq i \leq d_{n}^{j},\nonumber\\
\frac{L_{z(z')}}{2}|j,m,i\rangle_{z(z')}&=m |j,m,i\rangle_{z(z')},\nonumber\\
L_{+(+')}|j,m,i\rangle_{z(z')}&:=\frac{L_{x(x')}+i L_{y(y')}}{2}|j,m,i\rangle_{z(z')}=\sqrt{j(j+1)-m(m+1)} |j,m+1,i\rangle_{z(z')},\nonumber\\
L_{-(-')}|j,m,i\rangle_{z(z')}&:=\frac{L_{x(x')}-i L_{y(y')}}{2}|j,m,i\rangle_{z(z')}=\sqrt{j(j+1)-m(m-1)} |j,m-1,i\rangle_{z(z')},\nonumber\\
\frac{1}{4}(L_{x}^{2}+L_{y}^{2}+L_{z}^{2})|j,m,i\rangle_{z(z')}&=\frac{1}{4}(L_{x'}^{2}+L_{y'}^{2}+L_{z'}^{2})|j,m,i\rangle_{z(z')}=j(j+1) |j,m,i\rangle_{z(z')},\nonumber\\
L_{\alpha}&=\sum^{n}_{i=1}\sigma^{(i)}_{\alpha},~d^{j}_{n}=\frac{(2j+1)n!}{(n/2+j+1)!(n/2-j)!},
\end{align}
where $j_{\rm min}$ is 0 (1/2), and $j$, $m$ take (half-) integers for odd (even) $n$.
The index $i$ represents the number of ways of composing $n$ spins to obtain the total angular momentum $j$.
We refer to this basis as the irrep basis hereinafter.
In the definition, we introduce the $x'$ and $y'$ axes, which are orthogonal to Bob's $z'$ axis.
Although the choice of these axes has rotational ambiguity,
Bob can take any pair of these axes as the $x'$ and $y'$ axes because this does not affect 
the estimation of $\theta$.
Thus, we do not discuss the explicit direction of the $x'$ and $y'$ axes.
Note, also, that the operator $(L_{x}^{2}+L_{y}^{2}+L_{z}^{2})$ is invariant under the coordinate transformation $L_{x}^{2}+L_{y}^{2}+L_{z}^{2}=L_{x'}^{2}+L_{y'}^{2}+L_{z'}^{2}$.
In terms of the irrep basis, $|{\rm GHZ}\rangle$ is described as $|{\rm GHZ}\rangle=(|n/2,n/2,1\rangle_{z}+|n/2,-n/2,1\rangle_{z})/\sqrt{2}$.
For $j=n/2$, we simply represent the irrep basis $|n/2,m,1\rangle_{z'}$ ($d^{j}_{n}=1$ in this case) in terms of $|\uparrow (\downarrow)\rangle_{z'}$ as
\begin{equation}
|n/2,m,1\rangle_{z'}=\frac{1}{\sqrt{_{n}C_{m+n/2}}}(\underbrace{|\uparrow\uparrow\cdots\uparrow}_{m+n/2}\underbrace{\downarrow\cdots\downarrow}_{n/2-m}\rangle_{z'}+{\rm all~the~other~permutated~states}).
\label{eq:combination}
\end{equation}
The same representation also works for $|n/2,m,1\rangle_{z}$.
We emphasize that $|n/2,m,1\rangle_{z}$ still belongs to the $(j=n/2)$ subspace even when we expand this vector in terms of $|\uparrow (\downarrow)\rangle_{z'}$:  
\begin{equation}
|n/2,m,1\rangle_{z}=\sum_{-n/2\leq m\leq n/2}C_{m}|n/2,m,1\rangle_{z'}.
\label{eq:change}
\end{equation}
This expression is understood in terms of permutation symmetry.
We denote $U$ as the unitary matrix whose action is $U|\uparrow(\downarrow)\rangle_{z}=|\uparrow(\downarrow)\rangle_{z'}$.
According to Eq. (\ref{eq:combination}), the transformation between the irrep basis in both the $z$ and $z'$ representations is given as
\begin{equation}
|n/2,m,1\rangle_{z'}=U^{(1)}\cdots U^{(k)}\cdots U^{(n)}|n/2,m,1\rangle_{z}=\Bigl( \prod^{n}_{k=1}U^{(k)}\Bigr)|n/2,m,1\rangle_{z},
\end{equation}
where the index $k$ indicates that $U^{(k)}$ affects only the $k$-th qubit. Because $ \prod^{n}_{k=1}U^{(k)}$ is invariant under any permutation of the qubits, $|n/2,m,1\rangle_{z'}$ is symmetric under permutation (as $|n/2,m,1\rangle_{z}$ is symmetric).
Thus, $|n/2,m,1\rangle_{z'}$ is represented as the sum of $\{|n/2,m,1\rangle_{z'}\}_{m\leq n/2}$, as shown in Eq. (\ref{eq:change}).

For later convenience, we define the matrix elements $\overline{|j,m\rangle_{z'}\langle j,m'|}:=\frac{1}{d^{j}_{n}}\sum^{d^{j}_{n}}_{i=1}|j,m,i\rangle_{z'}\langle j,m',i|$, because none of the operations we address below depend on the index $i$.
An important point is that the expressions $\overline{|j,m\rangle_{z'}\langle j,m'|}$ are (super-) eigenvectors of the right-hand side of the Lindblad equation  (\ref{eq:lindbladsup}) with $\gamma'=0$, whose eigenvalues are given as
\begin{equation}
-i[\Omega L_{z'},\overline{|j,m\rangle_{z'}\langle j,m'|} ]+\gamma \Bigl( L_{z'}\overline{|j,m\rangle_{z'}\langle j,m'|}L_{z'}-\frac{1}{2}\{L_{z'}^2 \overline{|j,m\rangle_{z'}\langle j,m'|}\}\Bigr)=(-2 i\Omega(m-m') -2\gamma(m-m')^2)\overline{|j,m\rangle_{z'}\langle j,m'|}.
\label{eq:collective}
\end{equation}
We find that the initial state $\rho(0)=|{\rm GHZ}\rangle \langle {\rm GHZ}|$ can be rewritten using Eq. (\ref{eq:change}):
\begin{equation}
\rho(0)=\sum_{-n/2 \leq m,m'\leq n/2} \rho_{m,m'}\overline{|n/2,m\rangle_{z'}\langle n/2,m'|},
\end{equation}
where $\rho_{m,m'}={}_{z'}\langle n/2,m,1|{\rm GHZ}\rangle\langle {\rm GHZ}|n/2,m',1\rangle_{z'}$.
We can show that $\overline{|n/2,m\rangle_{z'}\langle n/2,m'|}=|n/2,m,1\rangle_{z'}\langle n/2,m',1|$.
The explicit form of $\rho_{m,m'}$ is
\begin{align}
\rho_{m,m'}=\frac{\sqrt{ _{n}C_{m+\frac{n}{2}}~_{n}C_{m'+\frac{n}{2}}}}{2^{n+1}}&\Bigl( (\sqrt{1+\cos \theta} )^{m+\frac{n}{2}}(-\sqrt{1-\cos \theta})^{\frac{n}{2}-m}+e^{-i n \phi}(\sqrt{1-\cos \theta})^{m+\frac{n}{2}}(\sqrt{1+\cos \theta})^{\frac{n}{2}-m}\Bigr)\times\nonumber\\
~\times&\Bigl( (\sqrt{1+\cos \theta} )^{m'+\frac{n}{2}}(-\sqrt{1-\cos \theta})^{\frac{n}{2}-m'}+e^{i n \phi}(\sqrt{1-\cos \theta})^{m'+\frac{n}{2}}(\sqrt{1+\cos \theta})^{\frac{n}{2}-m'}\Bigr).
\end{align}
Then the solution for $\gamma'=0$ is
\begin{equation}
\rho_{\gamma'=0}(t)=\sum_{-n/2 \leq m,m'\leq n/2}e^{-2 i \Omega(m-m')t - 2\gamma t (m-m')^2}\rho_{m,m'}\overline{|n/2,m\rangle_{z'} \langle n/2,m'|}.
\end{equation}

Next, we consider $\gamma'\neq 0$.
Because the independent dephasing term in Eq. (\ref{eq:lindbladsup}) commutes with the other two terms in the equation,
it is sufficient to consider its action independently.
The dynamical equation with only the third term is easily solved and thus the exact solution for $\gamma'\neq 0$ is written as follows:
\begin{align}
\rho(t)=&{\cal E}^{(n)}_{t}\cdots{\cal E}^{(i)}_{t}\cdots{\cal E}^{(1)}_{t}(\rho_{\gamma'=0}(t)),\nonumber\\
{\cal E}^{(i)}_{t}(\rho):=&\alpha(t)\rho+\beta(t)\sigma^{(i)}_{z'}\rho\sigma^{(i)}_{z'},
\end{align}
where $\alpha(t):=\bigl(\frac{1+e^{-2\gamma't}}{2}\bigr)$, and $\beta(t):=\bigl(\frac{1-e^{-2\gamma't}}{2}\bigr)$.
We rearrange the above equation as
\begin{align}
{\cal E}^{(n)}_{t}\cdots{\cal E}^{(i)}_{t}\cdots{\cal E}^{(1)}_{t}(\rho_{\gamma'=0}(t))=&\alpha^{n}(t)\rho_{\gamma'=0}(t) + \alpha^{n-1}(t)\beta(t)\sum_{i}\sigma_{z'}^{(i)}\rho_{\gamma'=0}(t)\sigma_{z'}^{(i)}\nonumber\\
&+\alpha^{n-2}(t)\beta^{2}(t)\sum_{1 \leq i < j \leq n}\sigma_{z'}^{(i)}\sigma_{z'}^{(j)}\rho_{\gamma'=0}(t)\sigma_{z}^{(j)}\sigma_{z'}^{(i)}+\cdots \nonumber\\
&+\cdots\beta^{n}(t)\sigma^{(1)}_{z'}\sigma^{(2)}_{z'}\cdots\sigma^{(n)}_{z'}\rho_{\gamma'=0}(t) \sigma^{(1)}_{z'}\sigma^{(2)}_{z'}\cdots\sigma^{(n)}_{z'}\nonumber\\
=&\sum^{n}_{k=0}\alpha^{n-k}(t)\beta^{k}(t)\hspace{-0.5cm}\sum_{1 \leq i_{1} < i_{2} < \cdots < i_{k} \leq n} \hspace{-0.5cm}\sigma^{(i_{1})}_{z'}\sigma^{(i_{2})}_{z'}\cdots\sigma^{(i_{k})}_{z'}\rho_{\gamma'=0}(t)  \sigma^{(i_{1})}_{z'}\sigma^{(i_{2})}_{z'}\cdots\sigma^{(i_{k})}_{z'},
\end{align}
where we assign $\alpha^{n}(t)\rho_{\gamma'=0}(t)$ as the $k=0$ term.
For convenience of notation,
we rewrite the above expression in terms of the irrep basis.
To this end,
we use the following: 
\begin{equation}
\sum_{1 \leq i_{1} < i_{2} < \cdots < i_{k} \leq n} \sigma^{(i_{1})}_{z'}\sigma^{(i_{2})}_{z'}\cdots\sigma^{(i_{k})}_{z'}\overline{|n/2,m\rangle_{z'} \langle n/2,m'|}  \sigma^{(i_{1})}_{z'}\sigma^{(i_{2})}_{z'}\cdots\sigma^{(i_{k})}_{z'}=\sum_{j}A^{(k)}_{j, m, m'}\overline{|j,m\rangle_{z'}\langle j,m'|}.
\label{eq:rewrite}
\end{equation}
The coefficients $A^{(k)}_{j, m, m'}$ are solved using recurrence relations.
To check this, we first use the following equality \cite{chase2008collective,baragiola2010collective}:
\begin{align}
\sum^{n}_{i=1}\sigma_{z'}^{(i)}\overline{|j,m\rangle_{z'}\langle j,m'|}&\sigma_{z'}^{(i)}\nonumber\\
=4\Bigl(a(n,j,m,m')&\overline{|j,m\rangle_{z'}\langle j,m'|}+b(n,j,m,m')\overline{|j-1,m\rangle_{z'}\langle j-1,m'|}+c(n,j,m,m')\overline{|j+1,m\rangle_{z'}\langle j+1,m'|}\Bigr),
\label{eq:spinbasis}
\end{align}
where
\begin{align}
a(n,j,m,m')=&m m' \frac{1}{2 j}\Bigl( 1+ \frac{(2j+1)\alpha^{j+1}_{n}}{(j+1)d^{j}_{n}}\Bigr),\nonumber\\
b(n,j,m,m')=&\sqrt{(j+m)(j-m)}\sqrt{(j+m')(j-m')}\frac{\alpha^{j}_{n}}{2 j d^{j}_{n}},\nonumber\\
c(n,j,m,m')=&\sqrt{(j+m+1)(j-m+1)}\sqrt{(j+m'+1)(j-m'+1)}\frac{\alpha^{j+1}_{n}}{2 (j+1) d^{j}_{n}},\nonumber\\
\alpha^{j}_{n}=&\sum^{n/2}_{j'=j}d^{j'}_{n}.
\end{align}
We evaluate the application of the following action to Eq. (\ref{eq:rewrite}):
\begin{equation}
\sum^{n}_{i=1}\sigma_{z'}^{(i)}\Bigl(\sum_{1 \leq i_{1} < i_{2} < \cdots < i_{k} \leq n} \sigma^{(i_{1})}_{z'}\sigma^{(i_{2})}_{z'}\cdots\sigma^{(i_{k})}_{z'}\overline{|n/2,m\rangle_{z'} \langle n/2,m'|}  \sigma^{(i_{1})}_{z'}\sigma^{(i_{2})}_{z'}\cdots\sigma^{(i_{k})}_{z'} \Bigr)\sigma_{z'}^{(i)}.
\end{equation}
By using $A^{k}_{j,m,m'}$, this expression can be written as
\begin{align}
\sum^{n}_{i=1}&\sigma_{z'}^{(i)}\Bigl(\sum_{1 \leq i_{1} < i_{2} < \cdots < i_{k} \leq n} \sigma^{(i_{1})}_{z'}\sigma^{(i_{2})}_{z'}\cdots\sigma^{(i_{k})}_{z'}\overline{|n/2,m\rangle_{z'} \langle n/2,m'|}  \sigma^{(i_{1})}_{z'}\sigma^{(i_{2})\cdots\sigma^{(i_{k})}_{z'}}_{z'}\Bigr)\sigma_{z'}^{(i)}\nonumber\\
&=\sum^{n}_{i=1}\sigma_{z'}^{(i)}\Bigl(\sum^{n/2}_{j=j_{\rm min}}A^{(k)}_{j, m, m'}\overline{|j,m\rangle_{z'}\langle j,m'|} \Bigr)\sigma_{z'}^{(i)}\nonumber\\
&=4 \sum^{n/2}_{j=j_{\rm min}}\bigl(a(n,j,m,m')A^{(k)}_{j, m, m'}+b(n,j+1,m,m')A^{(k)}_{j+1, m, m'}+c(n,j-1,m,m')A^{(k)}_{j-1, m, m'}\bigr)\overline{|j,m\rangle_{z'}\langle j,m'|},
\end{align}
where we use Eq. (\ref{eq:spinbasis}) and the conditions $b(n,n/2+1,m,m')=b(n,j_{\rm min},m,m')=0$ and $c(n,n/2,m,m')=c(n,j_{\rm min}-1,m,m')=0$ to align the summation range.
In addition, we can change the left-hand side of the above equation by a simple 
combinatorial calculation, as follows:
\begin{align}
\sum^{n}_{i=1}&\sigma_{z'}^{(i)}\Bigl(\sum_{1 \leq i_{1} < i_{2} < \cdots < i_{k} \leq n} \sigma^{(i_{1})}_{z'}\sigma^{(i_{2})}_{z'}\cdots\sigma^{(i_{k})}_{z'}\overline{|n/2,m\rangle_{z'} \langle n/2,m'|}  \sigma^{(i_{1})}_{z'}\sigma^{(i_{2})}_{z'}\cdots\sigma^{(i_{k})}_{z'} \Bigr)\sigma_{z'}^{(i)}\nonumber\\
=&(k+1)\sum_{1 \leq i_{1} < i_{2} < \cdots < i_{k+1} \leq n} \sigma^{(i_{1})}_{z'}\sigma^{(i_{2})}_{z'}\cdots\sigma^{(i_{k+1})}_{z'}\overline{|n/2,m\rangle_{z'} \langle n/2,m'|}  \sigma^{(i_{1})}_{z'}\sigma^{(i_{2})}_{z'}\cdots\sigma^{(i_{k+1})}_{z'}+\nonumber\\
&+(n-k+1)\sum_{1 \leq i_{1} < i_{2} < \cdots < i_{k-1} \leq n} \sigma^{(i_{1})}_{z'}\sigma^{(i_{2})}_{z'}\cdots\sigma^{(i_{k-1})}_{z'}\overline{|n/2,m\rangle_{z'} \langle n/2,m'|}  \sigma^{(i_{1})}_{z'}\sigma^{(i_{2})}_{z'}\cdots\sigma^{(i_{k-1})}_{z'}\nonumber\\
=&(k+1)\sum^{n/2}_{j=j_{\rm min}}A^{(k+1)}_{j, m, m'}\overline{|j,m\rangle_{z'}\langle j,m'|}+(n-k+1)\sum^{n/2}_{j=j_{\rm min}}A^{(k-1)}_{j, m, m'}\overline{|j,m\rangle_{z'}\langle j,m'|}.
\end{align}
By comparing the coefficients of each basis $\overline{|j,m\rangle_{z'}\langle j,m'|}$ in the above two equations, we obtain the recurrence relation of $A^{(k)}_{j, m, m'}$:
\begin{align}
A^{(k+1)}_{j, m, m'}=&\frac{1}{k+1}\Bigl(4a(n,j,m,m')A^{(k)}_{j, m, m'}+4b(n,j+1,m,m')A^{(k)}_{j+1, m, m'}\nonumber\\
&+4c(n,j-1,m,m')A^{(k)}_{j-1, m, m'}-(n-k+1)\sum^{n/2}_{j=j_{\rm min}}A^{(k-1)}_{j, m, m'}\Bigr).
\end{align}
When we use this recurrence relation, we assign the following conditions:
\begin{align}
\forall &m,m',~~A^{(0)}_{n/2, m, m'}=1,~A^{(0)}_{j\neq n/2, m, m'}=0,\nonumber\\
\forall j,&m,m',~~A^{(-1)}_{j, m, m'}=0,
\end{align}
where the first two conditions represent the initial conditions.
Thus, the exact solution of the dynamics (global magnetic field plus collective noise plus independent noise) is now written as 
\begin{equation}
\rho(t)=\sum^{n}_{k=0}\alpha^{n-k}(t)\beta^{k}(t) \sum^{n/2}_{j=j_{\rm min}}\sum_{-j \leq m,m'\leq j'}e^{-2 i \Omega (m-m')t-2\gamma t(m-m')^{2}}\rho_{m,m'}A^{(k)}_{j, m, m'} \overline{|j,m\rangle_{z'}\langle j, m'|}.
\label{eq:exact}
\end{equation}

\subsection{Analytical results for the scaling behavior}

Here we analytically evaluate the scaling behavior of the minimized uncertainty
\begin{equation}
\delta \theta^{\rm min}=\frac{\sqrt{P(t)(1-P(t))}}{|\frac{d P(t)}{d \theta}|\sqrt{M}}=\frac{\sqrt{P(t)(1-P(t))}}{|\frac{d P(t)}{d \theta}|\sqrt{T/t}},
\label{eq:var}
\end{equation} 
where $P(t)=\langle {\rm GHZ} |\rho (t) | {\rm GHZ}\rangle$ is the projection probability,
and $T$ is the total time allowed for the protocol.
According to Eq. (\ref{eq:exact}), the survival probability $P(t)$ is given as
\begin{align}
P(t)=&\langle {\rm GHZ} |\rho (t) | {\rm GHZ}\rangle\nonumber\\
=&\sum^{n}_{k=0}\alpha^{n-k}(t)\beta^{k}(t) \sum^{n/2}_{j=j_{\rm min}}\sum_{-j \leq m,m' \leq j}e^{- 2 i \Omega (m-m')t-2\gamma t(m-m')^{2}}\rho_{m,m'}A^{(k)}_{j, m, m'} \langle {\rm GHZ}\overline{|j,m\rangle_{z'}\langle j, m'|} {\rm GHZ}\rangle\nonumber\\
=&\sum^{n}_{k=0}\alpha^{n-k}(t)\beta^{k}(t) \sum_{-n/2\leq m,m'\leq n/2}e^{-i 2 \Omega (m-m')t-2\gamma t(m-m')^{2}}\rho_{m,m'}A^{(k)}_{n/2, m, m'} \langle {\rm GHZ}\overline{|n/2,m\rangle_{z'}\langle n/2, m'|} {\rm GHZ}\rangle\nonumber\\
=&\sum^{n}_{k=0}\alpha^{n-k}(t)\beta^{k}(t)\sum_{-n/2 \leq m,m' \leq n/2}e^{-2 i \Omega (m-m')t-2\gamma t(m-m')^{2}}
\rho_{m,m'} A^{(k)}_{n/2, m, m'} \langle {\rm GHZ}|n/2,m,1\rangle_{z'}\langle n/2, m',1| {\rm GHZ}\rangle\nonumber\\
=&\sum_{-n/2 \leq m,m' \leq n/2}e^{-2 i \Omega (m-m')t-2\gamma t(m-m')^{2}}\sum^{n}_{k=0}\alpha^{n-k}(t)\beta^{k}(t)A^{(k)}_{n/2, m, m'} B_{m}B_{m'},
\label{eq:exactprob}
\end{align}
where we define $B_{m}=|\langle{\rm GHZ}|n/2,m,1\rangle_{z'}|^{2}$ and use $\rho_{m,m'}={}_{z'}\langle n/2,m,1|{\rm GHZ}\rangle\langle {\rm GHZ}|n/2,m',1\rangle_{z'}$ and the fact that $|n/2,m,1\rangle_{z}$ still belongs to the $j=n/2$ subspace in the $z'$ bases. [See also Eq. (\ref{eq:change}).]
The explicit form of $B_{m}$ is given as
\begin{align}
B_{m}=&\frac{_{n}C_{m+\frac{n}{2}}}{2^{n+1}}\Bigl( (\sqrt{1+\cos \theta} )^{m+\frac{n}{2}}(-\sqrt{1-\cos \theta})^{\frac{n}{2}-m}+e^{-i n \phi}(\sqrt{1-\cos \theta})^{m+\frac{n}{2}}(\sqrt{1+\cos \theta} )^{\frac{n}{2}-m}\Bigr)\nonumber\\
&\times
\Bigl( (\sqrt{1+\cos \theta} )^{m+\frac{n}{2}}(-\sqrt{1-\cos \theta})^{\frac{n}{2}-m}+ e^{i n \phi}((\sqrt{1-\cos \theta})^{m+\frac{n}{2}}(\sqrt{1+\cos \theta} )^{\frac{n}{2}-m}\Bigr)\nonumber\\
=&\frac{_{n}C_{m+\frac{n}{2}}}{2^{n+1}}\Bigl(  (1+\cos \theta )^{m+\frac{n}{2}}(1-\cos \theta)^{\frac{n}{2}-m} + (1-\cos \theta)^{m+\frac{n}{2}}(1+\cos \theta)^{\frac{n}{2}-m} +(e^{i n \phi}+e^{-i n \phi}) \sin^{n} \theta (-1)^{\frac{n}{2}-m} \Bigr),
\label{eq:expa}
\end{align}
where we use $\sqrt{1-\cos^{2}\theta }=\sin \theta$ for $0\leq \theta \leq \pi$.
Assuming $n \Omega t,~n^{2} \gamma t,~n \gamma' t \ll 1$, we take the short time perturbation in Eq. (\ref{eq:exactprob}) up to the first order of $t$:
\begin{align}
P(t)\sim& \sum_{-n/2 \leq m,m'\leq n/2'}\bigl(1-2 i \Omega t (m-m') -2\gamma t (m-m')^{2}-\gamma' t(n A^{(0)}_{n/2, m, m'}+A^{(1)}_{n/2, m, m'})\bigr) B_{m}B_{m'}\nonumber\\
=&\sum_{-n/2 \leq m,m'\leq n/2'}\Bigl(1-2 i \Omega (m-m')t -2\gamma t(m-m')^{2}-\gamma' t \Bigl(n+\frac{4 m m'}{n} \Bigr)\Bigr) B_{m}B_{m'}\nonumber\\
=& \sum_{|m|\leq n/2}B_{m}\cdot\sum_{|m'|\leq n/2}B_{m'}-2 i \Omega t \Bigl(\sum_{ |m|\leq n/2} m B_{m}\cdot\sum_{|m'|\leq n/2}B_{m'}-\sum_{|m|\leq n/2} B_{m}\cdot\sum_{|m'|\leq n/2} m' B_{m'}  \Bigr)\nonumber\\
&-2\gamma t\Bigl( \sum_{|m|\leq n/2} m^{2} B_{m}\cdot\sum_{|m'|\leq n/2}B_{m'}-2 \sum_{|m|\leq n/2} m B_{m}\cdot\sum_{|m'|\leq n/2} m' B_{m'}+\sum_{|m|\leq n/2}B_{m}\cdot\sum_{|m'|\leq n/2}m'^{2}B_{m'}\Bigr)\nonumber\\
&-\gamma' t\Bigl( n \sum_{|m|\leq n/2}B_{m}\cdot\sum_{|m'|\leq n/2}B_{m'} +\frac{4}{n} \sum_{|m|\leq n/2} m B_{m}\cdot\sum_{|m'|\leq n/2} m' B_{m'} \Bigr)\nonumber\\\
=&\Bigl(\sum_{|m|\leq n/2}B_{m}\Bigr)^{2}-4 \gamma t\Bigl( \sum_{|m|\leq n/2} m^{2} B_{m}\cdot\sum_{|m|\leq n/2}B_{m}-\Bigl(\sum_{|m|\leq n/2} m B_{m}\Bigr)^{2}\Bigr)\nonumber\\
&-\gamma' t\Bigl(n \Bigl(\sum_{|m|\leq n/2}B_{m}\Bigr)^{2}+\frac{4}{n}\Bigl(\sum_{|m|\leq n/2} m B_{m} \Bigr)^{2}\Bigr).
\label{eq:probpert}
\end{align}

To evaluate this quantity, we have to evaluate only the following quantities:
\begin{equation}
\sum_{|m|\leq n/2}B_{m},~\sum_{|m|\leq n/2} m B_{m},~\sum_{|m|\leq n/2} m^{2} B_{m}.
\end{equation}
Note that the following 
formulae are satisfied:
\begin{align}
\sum^{n}_{m=0}~_{n}C_{m} X^{m}Y^{n-m}&=(X+Y)^{n},\nonumber\\
\sum^{n}_{m=0} m~ _{n}C_{m} X^{m}Y^{n-m}&=n X (X+Y)^{n-1},\nonumber\\
\sum^{n}_{m=0} m^{2}~ _{n}C_{m} X^{m}Y^{n-m}&=n X (X+Y)^{n-1}+(n^{2}-n)X^{2}(X+Y)^{n-2}.
\end{align}
We introduce
a new integer variable, $\mu:=m+n/2$, 
and 
let the sum range take integers.
We evaluate $\sum_{m} B_{m}$ as follows:
\begin{align}
\sum_{|m|\leq n/2}B_{m}=\sum^{n}_{\mu=0}B_{\mu-n/2}=&\sum^{n}_{\mu=0}\frac{_{n}C_{\mu}}{2^{n+1}}  (1+\cos \theta )^{\mu}(1-\cos \theta)^{n-\mu} + \sum^{n}_{\mu=0} \frac{_{n}C_{\mu}}{2^{n+1}} (1-\cos \theta)^{\mu}(1+\cos \theta )^{n-\mu}\nonumber\\
&+(e^{i n \phi}+e^{-i n \phi}) \sin^{n} \theta\sum^{n}_{\mu=0} \frac{_{n}C_{\mu}}{2^{n+1}}  (-1)^{n-\mu} \nonumber\\
=&\frac{2^{n}}{2^{n+1}}+\frac{2^{n}}{2^{n+1}} +( e^{i n \phi}+e^{-i n \phi}) \sin^{n} \theta \frac{(1-1)^{n}}{2^{n+1}}=1.
\end{align}
This equality is also understood in terms of the completeness of the basis $\{|n/2, m,1\rangle_{z'} \}$ in the $(j=n/2)$ subspace:
\begin{align}
\sum_{|m|\leq n/2}B_{m}=&\sum_{|m|\leq n/2}|\langle {\rm GHZ}|n/2,m,1\rangle_{z'}|^{2}=\sum_{|m|\leq n/2}\langle {\rm GHZ}|n/2,m,1\rangle_{z'}\langle n/2,m,1|{\rm GHZ}\rangle\nonumber\\
=&\langle {\rm GHZ}|\Bigl( \sum_{|m|\leq n/2}|n/2,m,1\rangle_{z'}\langle n/2,m,1| \Bigr)|{\rm GHZ}\rangle =\langle {\rm GHZ}|{\rm GHZ}\rangle=1.
\label{eq:zeroth}
\end{align}
Similarly, $\sum_{m} m B_{m}$ and $\sum_{m} m^{2} B_{m}$ are given by
\begin{align}
\sum_{|m|\leq n/2} m B_{m}=\sum^{n}_{\mu=0}\Bigl(\mu-\frac{n}{2}\Bigr) B_{\mu-n/2}=& \frac{n}{2^{n+1}} (1+\cos \theta )2^{n-1} +  \frac{n}{2^{n+1}}  (1-\cos \theta ) 2^{n-1}-\frac{n}{2}\sum^{n}_{\mu=0}B_{\mu-n/2}\nonumber\\
=&\frac{n}{4}\bigl((1+\cos \theta )+  (1-\cos \theta )\bigr)-\frac{n}{2}=0
\label{eq:first}
\end{align}
and
\begin{align}
\sum_{|m|\leq n/2} m^{2} B_{m}&=\sum^{n}_{\mu=0}\Bigl(\mu-\frac{n}{2}\Bigr)^2 B_{\mu-n/2}=\sum^{n}_{\mu=0} \mu^{2} B_{\mu-n/2}-n\sum^{n}_{\mu=0} \mu B_{\mu-n/2}+\frac{n^{2}}{4}\sum^{n}_{\mu=0} B_{\mu-n/2}\nonumber\\
=&\frac{n}{4}\bigl((1+\cos \theta )+ (1-\cos \theta)\bigr)+\frac{n^{2}-n}{8}\bigl((1+\cos \theta)^{2}+ (1-\cos \theta)^{2}\bigr)-\frac{n^{2}}{4}\nonumber\\
=&\frac{n}{2}+\frac{n^{2}-n}{4}(1+\cos^{2}\theta)-\frac{n^{2}}{4}=\frac{n^{2}}{4}\cos^{2} \theta+\frac{n}{4}(1-\cos^{2} \theta).
\end{align}
These equations, as well as Eq. (\ref{eq:zeroth}), are calculated as follows:
\begin{align}
\sum_{|m|\leq n/2} m B_{m}=&\sum_{|m|\leq n/2} m \langle {\rm GHZ}|n/2,m,1\rangle_{z'}\langle n/2,m,1|{\rm GHZ}\rangle=
\langle {\rm GHZ}|\sum_{|m|\leq n/2} \Bigl( m |n/2,m,1\rangle_{z'}\langle n/2,m,1| \Bigr)|{\rm GHZ}\rangle\nonumber\\
=& \langle {\rm GHZ}|\sum_{|m|\leq n/2} \Bigl( \frac{L_{z'}}{2} |n/2,m,1\rangle_{z'}\langle n/2,m,1| \Bigr)|{\rm GHZ}\rangle
= \langle {\rm GHZ}| \frac{L_{z'}}{2}|{\rm GHZ}\rangle\nonumber\\
=& \frac{1}{2}\langle {\rm GHZ}|(\cos \theta L_{z}+\cos\phi \sin \theta L_{x}+ i \sin\phi \sin \theta L_{y}) |{\rm GHZ}\rangle\nonumber\\
=&\frac{\cos \theta}{4}\Bigl(\langle \frac{n}{2},\frac{n}{2},1|L_{z}|\frac{n}{2},\frac{n}{2},1\rangle+\langle \frac{n}{2},-\frac{n}{2},1|L_{z}|\frac{n}{2},-\frac{n}{2},1\rangle\Bigr)=0
\end{align}
and
\begin{align}
\sum_{|m|\leq n/2} m^{2} B_{m}=&\langle {\rm GHZ}|\sum_{|m|\leq n/2} \Bigl( \frac{L^{2}_{z'}}{4} |n/2,m,1\rangle_{z'}\langle n/2,m,1| \Bigr)|{\rm GHZ}\rangle=\frac{1}{4}\langle {\rm GHZ}|L^{2}_{z'}|{\rm GHZ}\rangle\nonumber\\
=&\frac{1}{4}\langle {\rm GHZ}| (\cos \theta L_{z}+\cos\phi \sin \theta L_{x}+ i \sin\phi \sin \theta L_{y})^{2} |{\rm GHZ}\rangle\nonumber\\
=&\frac{1}{4}\langle {\rm GHZ}| (\cos^{2} \theta L^{2}_{z}+ \sin^{2} \theta (L_{+}L_{-}+L_{-}L_{+}) |{\rm GHZ}\rangle\nonumber\\
=&\frac{n^{2}}{4}\cos^{2} \theta + \frac{n}{4}(1-\cos^{2}\theta).
\end{align}

By using the above equations, Eq. (\ref{eq:probpert}) is finally evaluated as
\begin{equation}
P(t)=1-\gamma t (n^{2}\cos^{2} \theta+n(1-\cos^{2} \theta))
-\gamma' t n.
\label{eq:probscal}
\end{equation}
If we consider the estimation using collective dephasing ($\gamma \neq 0$) and assign the scaling behavior $t=t_{0}/n^{2}$ with a constant $t_{0}$, 
we find that $P(t)$ and $1-P(t)$ scale as ${\cal O}(n^{0})$.
In addition, $|d P(t)/d \theta |$ has the ${\cal O}(n^{0})$ dependence.
This result implies that the minimized uncertainty (\ref{eq:var}) scales as  
\begin{equation}
\delta \theta^{\rm min}=\frac{\sqrt{P(t)(1-P(t))}}{|\frac{d P(t)}{d \theta}|\sqrt{T/t}}=\frac{{\cal O}(n^{0})}{{\cal O}(n^{0})\sqrt{n^{2} T/t_{0}}}={\cal O}(n^{-1})
\end{equation}
for large $n$, which is the Heisenberg scaling.
Thus, by utilizing the short time perturbation, we show that the estimation of $\theta$ by collective dephasing
achieves the HL even under the effect of independent dephasing.
Note that $\theta$ can be estimated without knowing the value of $\phi$ in a short time regime because $P(t)$ is independent of $\phi$ in this regime.

In our scheme, we focus only on independent dephasing.
However, the above calculation also works for any type of independent noise.
According to the definition of the independence of noise, any independent noise behaves as $\sim \gamma' n t$ in a short time region, like the last term in Eq. (\ref{eq:probscal}).
Thus, if the $n^{2}$ term in Eq. (\ref{eq:probscal}) is present (or equivalently, if collective dephasing noise exists), we achieve the HL under any independent noise in the same manner as in the above discussion.
We should also point out that we cannot achieve the HL in the short time regime if we consider another type of collective dephasing noise with quadratic decay.
In this case, the calculation shown above reveals that the estimation of $\theta$ using the quadratic collective dephasing
achieves the SQL at best.

\subsection{Brief review of quantum Fisher information}

Here we briefly review the quantum Fisher information \cite{helstrom1976quantum}.
We focus only on single parameter estimation, where $\theta$ denotes the parameter to be estimated.
We have a density matrix $\rho_{\theta}$ in which the information on $\theta$ is imprinted and perform a POVM $\{ \Pi_{l}\}$ $(\Pi_{l}\geq 0, ~\sum_{l}\Pi_{l}={\mathbb I})$ on $\rho_{\theta}$.
From this measurement, we obtain a measurement outcome $l$ with a probability
\begin{equation}
P(l|\theta)={\rm Tr} \bigl( \Pi_{l}\rho_{\theta}\bigr).
\end{equation}
We prepare the state $\rho_{\theta}$ and perform POVM measurements with $\Pi_{l}$.
Suppose that we repeat these steps $M$ times.
Now we introduce an estimator $\tilde{\theta}(\vec{l})$, which is a function of $M$ outcomes $\vec{l}=\{l_{1},l_{2},l_{3},\cdots,l_{M}\}$, and identify the value of this estimator as the true value of the parameter $\theta$.
The precision of the estimation is determined by the uncertainty, $\delta \theta := \sqrt{\langle (\tilde\theta-\theta)^{2} \rangle}$, where the average ($\langle \bullet \rangle$) is defined as
\begin{equation}
\langle f \rangle=\sum_{\vec {l}}f(\vec{l})~\prod^{M}_{k=1}P(l_{k}|\theta)
\end{equation}
for a function of $M$ outcomes.
The following classical Cram\'er-Rao bound is satisfied for any estimator under the unbiased condition $\langle\tilde{\theta}\rangle=\theta$,
\begin{equation}
\delta \theta\geq 1/\sqrt{M F_{\theta}(\{\Pi_{l}\})},
\label{eq:cr}
\end{equation}
where $F_{\theta}(\{\Pi_{l}\})$ is the Fisher information, which is defined as
\begin{equation}
F_{\theta}(\{\Pi_{l}\})=\sum_{l} P(l|\theta)\frac{\partial \log P(l|\theta)}{\partial \theta}\frac{\log \partial P(l|\theta)}{\partial \theta}.
\label{eq:clasfi}
\end{equation}
In particular, a two-valued measurement $\{\Pi, {\mathbb I}-\Pi\}$ gives
\begin{equation}
F_{\theta}(\{\Pi,{\mathbb I}-\Pi \})=\frac{|d P(\theta)/d \theta|^{2}}{P(\theta)\bigl(1-P(\theta)\bigr)},
\end{equation}
where $P(\theta)={\rm Tr}(\Pi \rho_{\theta})$.

In quantum estimation, we can minimize the uncertainty $\delta \theta$ by choosing the best POVMs.
We have the following quantum Cram\'er-Rao bound for any POVM $\{\Pi_{l} \}$:
\begin{equation}
F_{\theta}(\{\Pi_{l}\})\leq F^{(Q)}_{\theta},
\label{eq:quanf}
\end{equation}
where $F^{(Q)}_{\theta}$ is called the quantum Fisher information and is defined as follows:
\begin{equation}
F^{(Q)}_{\theta}={\rm Tr}(L^{2}_{\theta}\rho_{\theta}),~~
\frac{\partial \rho_{\theta}}{\partial\theta}=\frac{1}{2}\{L_{\theta},\rho_{\theta}\}.
\end{equation}
By combining Eqs. (\ref{eq:cr}) and (\ref{eq:quanf}), we obtain a sequence of inequalities:
\begin{equation}
\delta \theta\geq 1/\sqrt{M F_{\theta}(\{\Pi_{l}\})}\geq 1/\sqrt{M F^{(Q)}_{\theta}}.
\end{equation}
For single-parameter estimation, it is shown that the second inequality can be saturated by taking an appropriate POVM, although that POVM may depend on the value of the parameter to be estimated.

\bibliography{metrology}

\begin{thebibliography}{70}%
\makeatletter
\providecommand \@ifxundefined [1]{%
 \@ifx{#1\undefined}
}%
\providecommand \@ifnum [1]{%
 \ifnum #1\expandafter \@firstoftwo
 \else \expandafter \@secondoftwo
 \fi
}%
\providecommand \@ifx [1]{%
 \ifx #1\expandafter \@firstoftwo
 \else \expandafter \@secondoftwo
 \fi
}%
\providecommand \natexlab [1]{#1}%
\providecommand \enquote  [1]{``#1''}%
\providecommand \bibnamefont  [1]{#1}%
\providecommand \bibfnamefont [1]{#1}%
\providecommand \citenamefont [1]{#1}%
\providecommand \href@noop [0]{\@secondoftwo}%
\providecommand \href [0]{\begingroup \@sanitize@url \@href}%
\providecommand \@href[1]{\@@startlink{#1}\@@href}%
\providecommand \@@href[1]{\endgroup#1\@@endlink}%
\providecommand \@sanitize@url [0]{\catcode `\\12\catcode `\$12\catcode
  `\&12\catcode `\#12\catcode `\^12\catcode `\_12\catcode `\%12\relax}%
\providecommand \@@startlink[1]{}%
\providecommand \@@endlink[0]{}%
\providecommand \url  [0]{\begingroup\@sanitize@url \@url }%
\providecommand \@url [1]{\endgroup\@href {#1}{\urlprefix }}%
\providecommand \urlprefix  [0]{URL }%
\providecommand \Eprint [0]{\href }%
\providecommand \doibase [0]{https://doi.org/}%
\providecommand \selectlanguage [0]{\@gobble}%
\providecommand \bibinfo  [0]{\@secondoftwo}%
\providecommand \bibfield  [0]{\@secondoftwo}%
\providecommand \translation [1]{[#1]}%
\providecommand \BibitemOpen [0]{}%
\providecommand \bibitemStop [0]{}%
\providecommand \bibitemNoStop [0]{.\EOS\space}%
\providecommand \EOS [0]{\spacefactor3000\relax}%
\providecommand \BibitemShut  [1]{\csname bibitem#1\endcsname}%
\let\auto@bib@innerbib\@empty
\bibitem [{\citenamefont {Huber}\ \emph {et~al.}(2008)\citenamefont {Huber},
  \citenamefont {Koshnick}, \citenamefont {Bluhm}, \citenamefont {Archuleta},
  \citenamefont {Azua}, \citenamefont {Bj{\"o}rnsson}, \citenamefont {Gardner},
  \citenamefont {Halloran}, \citenamefont {Lucero},\ and\ \citenamefont
  {Moler}}]{huber2008gradiometric}%
  \BibitemOpen
  \bibfield  {author} {\bibinfo {author} {\bibfnamefont {M.~E.}\ \bibnamefont
  {Huber}}, \bibinfo {author} {\bibfnamefont {N.~C.}\ \bibnamefont {Koshnick}},
  \bibinfo {author} {\bibfnamefont {H.}~\bibnamefont {Bluhm}}, \bibinfo
  {author} {\bibfnamefont {L.~J.}\ \bibnamefont {Archuleta}}, \bibinfo {author}
  {\bibfnamefont {T.}~\bibnamefont {Azua}}, \bibinfo {author} {\bibfnamefont
  {P.~G.}\ \bibnamefont {Bj{\"o}rnsson}}, \bibinfo {author} {\bibfnamefont
  {B.~W.}\ \bibnamefont {Gardner}}, \bibinfo {author} {\bibfnamefont {S.~T.}\
  \bibnamefont {Halloran}}, \bibinfo {author} {\bibfnamefont {E.~A.}\
  \bibnamefont {Lucero}},\ and\ \bibinfo {author} {\bibfnamefont {K.~A.}\
  \bibnamefont {Moler}},\ }\href@noop {} {\bibfield  {journal} {\bibinfo
  {journal} {Review of Scientific Instruments}\ }\textbf {\bibinfo {volume}
  {79}},\ \bibinfo {pages} {053704} (\bibinfo {year} {2008})}\BibitemShut
  {NoStop}%
\bibitem [{\citenamefont {Ramsden}(2011)}]{ramsden2011hall}%
  \BibitemOpen
  \bibfield  {author} {\bibinfo {author} {\bibfnamefont {E.}~\bibnamefont
  {Ramsden}},\ }\href@noop {} {\emph {\bibinfo {title} {Hall-effect sensors:
  theory and application}}}\ (\bibinfo  {publisher} {Elsevier},\ \bibinfo
  {year} {2011})\BibitemShut {NoStop}%
\bibitem [{\citenamefont {Poggio}\ and\ \citenamefont
  {Degen}(2010)}]{poggio2010force}%
  \BibitemOpen
  \bibfield  {author} {\bibinfo {author} {\bibfnamefont {M.}~\bibnamefont
  {Poggio}}\ and\ \bibinfo {author} {\bibfnamefont {C.~L.}\ \bibnamefont
  {Degen}},\ }\href@noop {} {\bibfield  {journal} {\bibinfo  {journal}
  {Nanotechnology}\ }\textbf {\bibinfo {volume} {21}},\ \bibinfo {pages}
  {342001} (\bibinfo {year} {2010})}\BibitemShut {NoStop}%
\bibitem [{\citenamefont {Helstrom}(1976)}]{helstrom1976quantum}%
  \BibitemOpen
  \bibfield  {author} {\bibinfo {author} {\bibfnamefont {C.~W.}\ \bibnamefont
  {Helstrom}},\ }\href@noop {} {\emph {\bibinfo {title} {Quantum detection and
  estimation theory}}},\ Vol.~\bibinfo {volume} {84}\ (\bibinfo  {publisher}
  {Academic press New York},\ \bibinfo {year} {1976})\BibitemShut {NoStop}%
\bibitem [{\citenamefont {Dunningham}(2006)}]{dunningham2006using}%
  \BibitemOpen
  \bibfield  {author} {\bibinfo {author} {\bibfnamefont {J.~A.}\ \bibnamefont
  {Dunningham}},\ }\href@noop {} {\bibfield  {journal} {\bibinfo  {journal}
  {Contemporary physics}\ }\textbf {\bibinfo {volume} {47}},\ \bibinfo {pages}
  {257} (\bibinfo {year} {2006})}\BibitemShut {NoStop}%
\bibitem [{\citenamefont {Holevo}(2011)}]{holevo2011probabilistic}%
  \BibitemOpen
  \bibfield  {author} {\bibinfo {author} {\bibfnamefont {A.~S.}\ \bibnamefont
  {Holevo}},\ }\href@noop {} {\emph {\bibinfo {title} {Probabilistic and
  statistical aspects of quantum theory}}},\ Vol.~\bibinfo {volume} {1}\
  (\bibinfo  {publisher} {Springer Science \& Business Media},\ \bibinfo {year}
  {2011})\BibitemShut {NoStop}%
\bibitem [{\citenamefont {Caves}(1981)}]{caves1981quantum}%
  \BibitemOpen
  \bibfield  {author} {\bibinfo {author} {\bibfnamefont {C.~M.}\ \bibnamefont
  {Caves}},\ }\href@noop {} {\bibfield  {journal} {\bibinfo  {journal}
  {Physical Review D}\ }\textbf {\bibinfo {volume} {23}},\ \bibinfo {pages}
  {1693} (\bibinfo {year} {1981})}\BibitemShut {NoStop}%
\bibitem [{\citenamefont {Giovannetti}\ \emph {et~al.}(2004)\citenamefont
  {Giovannetti}, \citenamefont {Lloyd},\ and\ \citenamefont
  {Maccone}}]{giovannetti2004quantum}%
  \BibitemOpen
  \bibfield  {author} {\bibinfo {author} {\bibfnamefont {V.}~\bibnamefont
  {Giovannetti}}, \bibinfo {author} {\bibfnamefont {S.}~\bibnamefont {Lloyd}},\
  and\ \bibinfo {author} {\bibfnamefont {L.}~\bibnamefont {Maccone}},\
  }\href@noop {} {\bibfield  {journal} {\bibinfo  {journal} {Science}\ }\textbf
  {\bibinfo {volume} {306}},\ \bibinfo {pages} {1330} (\bibinfo {year}
  {2004})}\BibitemShut {NoStop}%
\bibitem [{\citenamefont {Giovannetti}\ \emph {et~al.}(2006)\citenamefont
  {Giovannetti}, \citenamefont {Lloyd},\ and\ \citenamefont
  {Maccone}}]{giovannetti2006quantum}%
  \BibitemOpen
  \bibfield  {author} {\bibinfo {author} {\bibfnamefont {V.}~\bibnamefont
  {Giovannetti}}, \bibinfo {author} {\bibfnamefont {S.}~\bibnamefont {Lloyd}},\
  and\ \bibinfo {author} {\bibfnamefont {L.}~\bibnamefont {Maccone}},\
  }\href@noop {} {\bibfield  {journal} {\bibinfo  {journal} {Physical review
  letters}\ }\textbf {\bibinfo {volume} {96}},\ \bibinfo {pages} {010401}
  (\bibinfo {year} {2006})}\BibitemShut {NoStop}%
\bibitem [{\citenamefont {Simon}\ \emph {et~al.}(2017)\citenamefont {Simon},
  \citenamefont {Jaeger},\ and\ \citenamefont {Sergienko}}]{simon2017quantum}%
  \BibitemOpen
  \bibfield  {author} {\bibinfo {author} {\bibfnamefont {D.~S.}\ \bibnamefont
  {Simon}}, \bibinfo {author} {\bibfnamefont {G.}~\bibnamefont {Jaeger}},\ and\
  \bibinfo {author} {\bibfnamefont {A.~V.}\ \bibnamefont {Sergienko}},\ }in\
  \href@noop {} {\emph {\bibinfo {booktitle} {Quantum Metrology, Imaging, and
  Communication}}}\ (\bibinfo  {publisher} {Springer},\ \bibinfo {year}
  {2017})\ pp.\ \bibinfo {pages} {91--112}\BibitemShut {NoStop}%
\bibitem [{\citenamefont {Giovannetti}\ \emph {et~al.}(2011)\citenamefont
  {Giovannetti}, \citenamefont {Lloyd},\ and\ \citenamefont
  {Maccone}}]{giovannetti2011advances}%
  \BibitemOpen
  \bibfield  {author} {\bibinfo {author} {\bibfnamefont {V.}~\bibnamefont
  {Giovannetti}}, \bibinfo {author} {\bibfnamefont {S.}~\bibnamefont {Lloyd}},\
  and\ \bibinfo {author} {\bibfnamefont {L.}~\bibnamefont {Maccone}},\
  }\href@noop {} {\bibfield  {journal} {\bibinfo  {journal} {Nature photonics}\
  }\textbf {\bibinfo {volume} {5}},\ \bibinfo {pages} {222} (\bibinfo {year}
  {2011})}\BibitemShut {NoStop}%
\bibitem [{\citenamefont {Taylor}\ and\ \citenamefont
  {Bowen}(2016)}]{taylor2016quantum}%
  \BibitemOpen
  \bibfield  {author} {\bibinfo {author} {\bibfnamefont {M.~A.}\ \bibnamefont
  {Taylor}}\ and\ \bibinfo {author} {\bibfnamefont {W.~P.}\ \bibnamefont
  {Bowen}},\ }\href@noop {} {\bibfield  {journal} {\bibinfo  {journal} {Physics
  Reports}\ }\textbf {\bibinfo {volume} {615}},\ \bibinfo {pages} {1} (\bibinfo
  {year} {2016})}\BibitemShut {NoStop}%
\bibitem [{\citenamefont {Degen}\ \emph {et~al.}(2017)\citenamefont {Degen},
  \citenamefont {Reinhard},\ and\ \citenamefont
  {Cappellaro}}]{degen2017quantum}%
  \BibitemOpen
  \bibfield  {author} {\bibinfo {author} {\bibfnamefont {C.~L.}\ \bibnamefont
  {Degen}}, \bibinfo {author} {\bibfnamefont {F.}~\bibnamefont {Reinhard}},\
  and\ \bibinfo {author} {\bibfnamefont {P.}~\bibnamefont {Cappellaro}},\
  }\href@noop {} {\bibfield  {journal} {\bibinfo  {journal} {Reviews of modern
  physics}\ }\textbf {\bibinfo {volume} {89}},\ \bibinfo {pages} {035002}
  (\bibinfo {year} {2017})}\BibitemShut {NoStop}%
\bibitem [{\citenamefont {Paris}(2009)}]{paris2009quantum}%
  \BibitemOpen
  \bibfield  {author} {\bibinfo {author} {\bibfnamefont {M.~G.}\ \bibnamefont
  {Paris}},\ }\href@noop {} {\bibfield  {journal} {\bibinfo  {journal}
  {International Journal of Quantum Information}\ }\textbf {\bibinfo {volume}
  {7}},\ \bibinfo {pages} {125} (\bibinfo {year} {2009})}\BibitemShut {NoStop}%
\bibitem [{\citenamefont {Wineland}\ \emph {et~al.}(1992)\citenamefont
  {Wineland}, \citenamefont {Bollinger}, \citenamefont {Itano}, \citenamefont
  {Moore},\ and\ \citenamefont {Heinzen}}]{wineland1992spin}%
  \BibitemOpen
  \bibfield  {author} {\bibinfo {author} {\bibfnamefont {D.~J.}\ \bibnamefont
  {Wineland}}, \bibinfo {author} {\bibfnamefont {J.~J.}\ \bibnamefont
  {Bollinger}}, \bibinfo {author} {\bibfnamefont {W.~M.}\ \bibnamefont
  {Itano}}, \bibinfo {author} {\bibfnamefont {F.}~\bibnamefont {Moore}},\ and\
  \bibinfo {author} {\bibfnamefont {D.}~\bibnamefont {Heinzen}},\ }\href@noop
  {} {\bibfield  {journal} {\bibinfo  {journal} {Physical Review A}\ }\textbf
  {\bibinfo {volume} {46}},\ \bibinfo {pages} {R6797} (\bibinfo {year}
  {1992})}\BibitemShut {NoStop}%
\bibitem [{\citenamefont {Wineland}\ \emph {et~al.}(1994)\citenamefont
  {Wineland}, \citenamefont {Bollinger}, \citenamefont {Itano},\ and\
  \citenamefont {Heinzen}}]{wineland1994squeezed}%
  \BibitemOpen
  \bibfield  {author} {\bibinfo {author} {\bibfnamefont {D.~J.}\ \bibnamefont
  {Wineland}}, \bibinfo {author} {\bibfnamefont {J.~J.}\ \bibnamefont
  {Bollinger}}, \bibinfo {author} {\bibfnamefont {W.~M.}\ \bibnamefont
  {Itano}},\ and\ \bibinfo {author} {\bibfnamefont {D.}~\bibnamefont
  {Heinzen}},\ }\href@noop {} {\bibfield  {journal} {\bibinfo  {journal}
  {Physical Review A}\ }\textbf {\bibinfo {volume} {50}},\ \bibinfo {pages}
  {67} (\bibinfo {year} {1994})}\BibitemShut {NoStop}%
\bibitem [{\citenamefont {T{\'o}th}\ and\ \citenamefont
  {Apellaniz}(2014)}]{toth2014quantum}%
  \BibitemOpen
  \bibfield  {author} {\bibinfo {author} {\bibfnamefont {G.}~\bibnamefont
  {T{\'o}th}}\ and\ \bibinfo {author} {\bibfnamefont {I.}~\bibnamefont
  {Apellaniz}},\ }\href@noop {} {\bibfield  {journal} {\bibinfo  {journal}
  {Journal of Physics A: Mathematical and Theoretical}\ }\textbf {\bibinfo
  {volume} {47}},\ \bibinfo {pages} {424006} (\bibinfo {year}
  {2014})}\BibitemShut {NoStop}%
\bibitem [{\citenamefont {Bollinger}\ \emph {et~al.}(1996)\citenamefont
  {Bollinger}, \citenamefont {Itano}, \citenamefont {Wineland},\ and\
  \citenamefont {Heinzen}}]{bollinger1996optimal}%
  \BibitemOpen
  \bibfield  {author} {\bibinfo {author} {\bibfnamefont {J.~J.}\ \bibnamefont
  {Bollinger}}, \bibinfo {author} {\bibfnamefont {W.~M.}\ \bibnamefont
  {Itano}}, \bibinfo {author} {\bibfnamefont {D.~J.}\ \bibnamefont
  {Wineland}},\ and\ \bibinfo {author} {\bibfnamefont {D.~J.}\ \bibnamefont
  {Heinzen}},\ }\href@noop {} {\bibfield  {journal} {\bibinfo  {journal}
  {Physical Review A}\ }\textbf {\bibinfo {volume} {54}},\ \bibinfo {pages}
  {R4649} (\bibinfo {year} {1996})}\BibitemShut {NoStop}%
\bibitem [{\citenamefont {Macieszczak}(2015)}]{macieszczak2015zeno}%
  \BibitemOpen
  \bibfield  {author} {\bibinfo {author} {\bibfnamefont {K.}~\bibnamefont
  {Macieszczak}},\ }\href@noop {} {\bibfield  {journal} {\bibinfo  {journal}
  {Physical Review A}\ }\textbf {\bibinfo {volume} {92}},\ \bibinfo {pages}
  {010102} (\bibinfo {year} {2015})}\BibitemShut {NoStop}%
\bibitem [{\citenamefont {Huelga}\ \emph {et~al.}(1997)\citenamefont {Huelga},
  \citenamefont {Macchiavello}, \citenamefont {Pellizzari}, \citenamefont
  {Ekert}, \citenamefont {Plenio},\ and\ \citenamefont
  {Cirac}}]{PhysRevLett.79.3865}%
  \BibitemOpen
  \bibfield  {author} {\bibinfo {author} {\bibfnamefont {S.~F.}\ \bibnamefont
  {Huelga}}, \bibinfo {author} {\bibfnamefont {C.}~\bibnamefont
  {Macchiavello}}, \bibinfo {author} {\bibfnamefont {T.}~\bibnamefont
  {Pellizzari}}, \bibinfo {author} {\bibfnamefont {A.~K.}\ \bibnamefont
  {Ekert}}, \bibinfo {author} {\bibfnamefont {M.~B.}\ \bibnamefont {Plenio}},\
  and\ \bibinfo {author} {\bibfnamefont {J.~I.}\ \bibnamefont {Cirac}},\ }\href
  {https://doi.org/10.1103/PhysRevLett.79.3865} {\bibfield  {journal} {\bibinfo
   {journal} {Phys. Rev. Lett.}\ }\textbf {\bibinfo {volume} {79}},\ \bibinfo
  {pages} {3865} (\bibinfo {year} {1997})}\BibitemShut {NoStop}%
\bibitem [{\citenamefont {G{\'o}recka}\ \emph {et~al.}(2018)\citenamefont
  {G{\'o}recka}, \citenamefont {Pollock}, \citenamefont {Liuzzo-Scorpo},
  \citenamefont {Nichols}, \citenamefont {Adesso},\ and\ \citenamefont
  {Modi}}]{gorecka2018noisy}%
  \BibitemOpen
  \bibfield  {author} {\bibinfo {author} {\bibfnamefont {A.}~\bibnamefont
  {G{\'o}recka}}, \bibinfo {author} {\bibfnamefont {F.~A.}\ \bibnamefont
  {Pollock}}, \bibinfo {author} {\bibfnamefont {P.}~\bibnamefont
  {Liuzzo-Scorpo}}, \bibinfo {author} {\bibfnamefont {R.}~\bibnamefont
  {Nichols}}, \bibinfo {author} {\bibfnamefont {G.}~\bibnamefont {Adesso}},\
  and\ \bibinfo {author} {\bibfnamefont {K.}~\bibnamefont {Modi}},\ }\href@noop
  {} {\bibfield  {journal} {\bibinfo  {journal} {New Journal of Physics}\
  }\textbf {\bibinfo {volume} {20}},\ \bibinfo {pages} {083008} (\bibinfo
  {year} {2018})}\BibitemShut {NoStop}%
\bibitem [{\citenamefont {Rossi}\ \emph {et~al.}(2020)\citenamefont {Rossi},
  \citenamefont {Albarelli}, \citenamefont {Tamascelli},\ and\ \citenamefont
  {Genoni}}]{rossi2020noisy}%
  \BibitemOpen
  \bibfield  {author} {\bibinfo {author} {\bibfnamefont {M.~A.}\ \bibnamefont
  {Rossi}}, \bibinfo {author} {\bibfnamefont {F.}~\bibnamefont {Albarelli}},
  \bibinfo {author} {\bibfnamefont {D.}~\bibnamefont {Tamascelli}},\ and\
  \bibinfo {author} {\bibfnamefont {M.~G.}\ \bibnamefont {Genoni}},\
  }\href@noop {} {\bibfield  {journal} {\bibinfo  {journal} {Physical Review
  Letters}\ }\textbf {\bibinfo {volume} {125}},\ \bibinfo {pages} {200505}
  (\bibinfo {year} {2020})}\BibitemShut {NoStop}%
\bibitem [{\citenamefont {Altenburg}\ \emph {et~al.}(2017)\citenamefont
  {Altenburg}, \citenamefont {Oszmaniec}, \citenamefont {W{\"o}lk},\ and\
  \citenamefont {G{\"u}hne}}]{altenburg2017estimation}%
  \BibitemOpen
  \bibfield  {author} {\bibinfo {author} {\bibfnamefont {S.}~\bibnamefont
  {Altenburg}}, \bibinfo {author} {\bibfnamefont {M.}~\bibnamefont
  {Oszmaniec}}, \bibinfo {author} {\bibfnamefont {S.}~\bibnamefont
  {W{\"o}lk}},\ and\ \bibinfo {author} {\bibfnamefont {O.}~\bibnamefont
  {G{\"u}hne}},\ }\href@noop {} {\bibfield  {journal} {\bibinfo  {journal}
  {Physical Review A}\ }\textbf {\bibinfo {volume} {96}},\ \bibinfo {pages}
  {042319} (\bibinfo {year} {2017})}\BibitemShut {NoStop}%
\bibitem [{\citenamefont {Schmitt}\ \emph {et~al.}(2017)\citenamefont
  {Schmitt}, \citenamefont {Gefen}, \citenamefont {St{\"u}rner}, \citenamefont
  {Unden}, \citenamefont {Wolff}, \citenamefont {M{\"u}ller}, \citenamefont
  {Scheuer}, \citenamefont {Naydenov}, \citenamefont {Markham}, \citenamefont
  {Pezzagna} \emph {et~al.}}]{schmitt2017submillihertz}%
  \BibitemOpen
  \bibfield  {author} {\bibinfo {author} {\bibfnamefont {S.}~\bibnamefont
  {Schmitt}}, \bibinfo {author} {\bibfnamefont {T.}~\bibnamefont {Gefen}},
  \bibinfo {author} {\bibfnamefont {F.~M.}\ \bibnamefont {St{\"u}rner}},
  \bibinfo {author} {\bibfnamefont {T.}~\bibnamefont {Unden}}, \bibinfo
  {author} {\bibfnamefont {G.}~\bibnamefont {Wolff}}, \bibinfo {author}
  {\bibfnamefont {C.}~\bibnamefont {M{\"u}ller}}, \bibinfo {author}
  {\bibfnamefont {J.}~\bibnamefont {Scheuer}}, \bibinfo {author} {\bibfnamefont
  {B.}~\bibnamefont {Naydenov}}, \bibinfo {author} {\bibfnamefont
  {M.}~\bibnamefont {Markham}}, \bibinfo {author} {\bibfnamefont
  {S.}~\bibnamefont {Pezzagna}}, \emph {et~al.},\ }\href@noop {} {\bibfield
  {journal} {\bibinfo  {journal} {Science}\ }\textbf {\bibinfo {volume}
  {356}},\ \bibinfo {pages} {832} (\bibinfo {year} {2017})}\BibitemShut
  {NoStop}%
\bibitem [{\citenamefont {Ledbetter}\ \emph {et~al.}(2012)\citenamefont
  {Ledbetter}, \citenamefont {Jensen}, \citenamefont {Fischer}, \citenamefont
  {Jarmola},\ and\ \citenamefont {Budker}}]{ledbetter2012gyroscopes}%
  \BibitemOpen
  \bibfield  {author} {\bibinfo {author} {\bibfnamefont {M.}~\bibnamefont
  {Ledbetter}}, \bibinfo {author} {\bibfnamefont {K.}~\bibnamefont {Jensen}},
  \bibinfo {author} {\bibfnamefont {R.}~\bibnamefont {Fischer}}, \bibinfo
  {author} {\bibfnamefont {A.}~\bibnamefont {Jarmola}},\ and\ \bibinfo {author}
  {\bibfnamefont {D.}~\bibnamefont {Budker}},\ }\href@noop {} {\bibfield
  {journal} {\bibinfo  {journal} {Physical Review A}\ }\textbf {\bibinfo
  {volume} {86}},\ \bibinfo {pages} {052116} (\bibinfo {year}
  {2012})}\BibitemShut {NoStop}%
\bibitem [{\citenamefont {Ajoy}\ and\ \citenamefont
  {Cappellaro}(2012)}]{ajoy2012stable}%
  \BibitemOpen
  \bibfield  {author} {\bibinfo {author} {\bibfnamefont {A.}~\bibnamefont
  {Ajoy}}\ and\ \bibinfo {author} {\bibfnamefont {P.}~\bibnamefont
  {Cappellaro}},\ }\href@noop {} {\bibfield  {journal} {\bibinfo  {journal}
  {Physical Review A}\ }\textbf {\bibinfo {volume} {86}},\ \bibinfo {pages}
  {062104} (\bibinfo {year} {2012})}\BibitemShut {NoStop}%
\bibitem [{\citenamefont {Monz}\ \emph {et~al.}(2011)\citenamefont {Monz},
  \citenamefont {Schindler}, \citenamefont {Barreiro}, \citenamefont {Chwalla},
  \citenamefont {Nigg}, \citenamefont {Coish}, \citenamefont {Harlander},
  \citenamefont {H{\"a}nsel}, \citenamefont {Hennrich},\ and\ \citenamefont
  {Blatt}}]{monz201114}%
  \BibitemOpen
  \bibfield  {author} {\bibinfo {author} {\bibfnamefont {T.}~\bibnamefont
  {Monz}}, \bibinfo {author} {\bibfnamefont {P.}~\bibnamefont {Schindler}},
  \bibinfo {author} {\bibfnamefont {J.~T.}\ \bibnamefont {Barreiro}}, \bibinfo
  {author} {\bibfnamefont {M.}~\bibnamefont {Chwalla}}, \bibinfo {author}
  {\bibfnamefont {D.}~\bibnamefont {Nigg}}, \bibinfo {author} {\bibfnamefont
  {W.~A.}\ \bibnamefont {Coish}}, \bibinfo {author} {\bibfnamefont
  {M.}~\bibnamefont {Harlander}}, \bibinfo {author} {\bibfnamefont
  {W.}~\bibnamefont {H{\"a}nsel}}, \bibinfo {author} {\bibfnamefont
  {M.}~\bibnamefont {Hennrich}},\ and\ \bibinfo {author} {\bibfnamefont
  {R.}~\bibnamefont {Blatt}},\ }\href@noop {} {\bibfield  {journal} {\bibinfo
  {journal} {Physical Review Letters}\ }\textbf {\bibinfo {volume} {106}},\
  \bibinfo {pages} {130506} (\bibinfo {year} {2011})}\BibitemShut {NoStop}%
\bibitem [{\citenamefont {Greenberger}\ \emph {et~al.}(1990)\citenamefont
  {Greenberger}, \citenamefont {Horne}, \citenamefont {Shimony},\ and\
  \citenamefont {Zeilinger}}]{greenberger1990bell}%
  \BibitemOpen
  \bibfield  {author} {\bibinfo {author} {\bibfnamefont {D.~M.}\ \bibnamefont
  {Greenberger}}, \bibinfo {author} {\bibfnamefont {M.~A.}\ \bibnamefont
  {Horne}}, \bibinfo {author} {\bibfnamefont {A.}~\bibnamefont {Shimony}},\
  and\ \bibinfo {author} {\bibfnamefont {A.}~\bibnamefont {Zeilinger}},\
  }\href@noop {} {\bibfield  {journal} {\bibinfo  {journal} {American Journal
  of Physics}\ }\textbf {\bibinfo {volume} {58}},\ \bibinfo {pages} {1131}
  (\bibinfo {year} {1990})}\BibitemShut {NoStop}%
\bibitem [{\citenamefont {DiCarlo}\ \emph {et~al.}(2010)\citenamefont
  {DiCarlo}, \citenamefont {Reed}, \citenamefont {Sun}, \citenamefont
  {Johnson}, \citenamefont {Chow}, \citenamefont {Gambetta}, \citenamefont
  {Frunzio}, \citenamefont {Girvin}, \citenamefont {Devoret},\ and\
  \citenamefont {Schoelkopf}}]{dicarlo2010preparation}%
  \BibitemOpen
  \bibfield  {author} {\bibinfo {author} {\bibfnamefont {L.}~\bibnamefont
  {DiCarlo}}, \bibinfo {author} {\bibfnamefont {M.~D.}\ \bibnamefont {Reed}},
  \bibinfo {author} {\bibfnamefont {L.}~\bibnamefont {Sun}}, \bibinfo {author}
  {\bibfnamefont {B.~R.}\ \bibnamefont {Johnson}}, \bibinfo {author}
  {\bibfnamefont {J.~M.}\ \bibnamefont {Chow}}, \bibinfo {author}
  {\bibfnamefont {J.~M.}\ \bibnamefont {Gambetta}}, \bibinfo {author}
  {\bibfnamefont {L.}~\bibnamefont {Frunzio}}, \bibinfo {author} {\bibfnamefont
  {S.~M.}\ \bibnamefont {Girvin}}, \bibinfo {author} {\bibfnamefont {M.~H.}\
  \bibnamefont {Devoret}},\ and\ \bibinfo {author} {\bibfnamefont {R.~J.}\
  \bibnamefont {Schoelkopf}},\ }\href@noop {} {\bibfield  {journal} {\bibinfo
  {journal} {Nature}\ }\textbf {\bibinfo {volume} {467}},\ \bibinfo {pages}
  {574} (\bibinfo {year} {2010})}\BibitemShut {NoStop}%
\bibitem [{\citenamefont {Holland}\ and\ \citenamefont
  {Burnett}(1993)}]{holland1993interferometric}%
  \BibitemOpen
  \bibfield  {author} {\bibinfo {author} {\bibfnamefont {M.}~\bibnamefont
  {Holland}}\ and\ \bibinfo {author} {\bibfnamefont {K.}~\bibnamefont
  {Burnett}},\ }\href@noop {} {\bibfield  {journal} {\bibinfo  {journal}
  {Physical review letters}\ }\textbf {\bibinfo {volume} {71}},\ \bibinfo
  {pages} {1355} (\bibinfo {year} {1993})}\BibitemShut {NoStop}%
\bibitem [{\citenamefont {Nagata}\ \emph {et~al.}(2007)\citenamefont {Nagata},
  \citenamefont {Okamoto}, \citenamefont {O'brien}, \citenamefont {Sasaki},\
  and\ \citenamefont {Takeuchi}}]{nagata2007beating}%
  \BibitemOpen
  \bibfield  {author} {\bibinfo {author} {\bibfnamefont {T.}~\bibnamefont
  {Nagata}}, \bibinfo {author} {\bibfnamefont {R.}~\bibnamefont {Okamoto}},
  \bibinfo {author} {\bibfnamefont {J.~L.}\ \bibnamefont {O'brien}}, \bibinfo
  {author} {\bibfnamefont {K.}~\bibnamefont {Sasaki}},\ and\ \bibinfo {author}
  {\bibfnamefont {S.}~\bibnamefont {Takeuchi}},\ }\href@noop {} {\bibfield
  {journal} {\bibinfo  {journal} {Science}\ }\textbf {\bibinfo {volume}
  {316}},\ \bibinfo {pages} {726} (\bibinfo {year} {2007})}\BibitemShut
  {NoStop}%
\bibitem [{\citenamefont {Jones}\ \emph {et~al.}(2009)\citenamefont {Jones},
  \citenamefont {Karlen}, \citenamefont {Fitzsimons}, \citenamefont {Ardavan},
  \citenamefont {Benjamin}, \citenamefont {Briggs},\ and\ \citenamefont
  {Morton}}]{jones2009magnetic}%
  \BibitemOpen
  \bibfield  {author} {\bibinfo {author} {\bibfnamefont {J.~A.}\ \bibnamefont
  {Jones}}, \bibinfo {author} {\bibfnamefont {S.~D.}\ \bibnamefont {Karlen}},
  \bibinfo {author} {\bibfnamefont {J.}~\bibnamefont {Fitzsimons}}, \bibinfo
  {author} {\bibfnamefont {A.}~\bibnamefont {Ardavan}}, \bibinfo {author}
  {\bibfnamefont {S.~C.}\ \bibnamefont {Benjamin}}, \bibinfo {author}
  {\bibfnamefont {G.~A.~D.}\ \bibnamefont {Briggs}},\ and\ \bibinfo {author}
  {\bibfnamefont {J.~J.}\ \bibnamefont {Morton}},\ }\href@noop {} {\bibfield
  {journal} {\bibinfo  {journal} {Science}\ }\textbf {\bibinfo {volume}
  {324}},\ \bibinfo {pages} {1166} (\bibinfo {year} {2009})}\BibitemShut
  {NoStop}%
\bibitem [{\citenamefont {Facon}\ \emph {et~al.}(2016)\citenamefont {Facon},
  \citenamefont {Dietsche}, \citenamefont {Grosso}, \citenamefont {Haroche},
  \citenamefont {Raimond}, \citenamefont {Brune},\ and\ \citenamefont
  {Gleyzes}}]{facon2016sensitive}%
  \BibitemOpen
  \bibfield  {author} {\bibinfo {author} {\bibfnamefont {A.}~\bibnamefont
  {Facon}}, \bibinfo {author} {\bibfnamefont {E.-K.}\ \bibnamefont {Dietsche}},
  \bibinfo {author} {\bibfnamefont {D.}~\bibnamefont {Grosso}}, \bibinfo
  {author} {\bibfnamefont {S.}~\bibnamefont {Haroche}}, \bibinfo {author}
  {\bibfnamefont {J.-M.}\ \bibnamefont {Raimond}}, \bibinfo {author}
  {\bibfnamefont {M.}~\bibnamefont {Brune}},\ and\ \bibinfo {author}
  {\bibfnamefont {S.}~\bibnamefont {Gleyzes}},\ }\href@noop {} {\bibfield
  {journal} {\bibinfo  {journal} {Nature}\ }\textbf {\bibinfo {volume} {535}},\
  \bibinfo {pages} {262} (\bibinfo {year} {2016})}\BibitemShut {NoStop}%
\bibitem [{\citenamefont {Kruse}\ \emph {et~al.}(2016)\citenamefont {Kruse},
  \citenamefont {Lange}, \citenamefont {Peise}, \citenamefont {L{\"u}cke},
  \citenamefont {Pezz{\`e}}, \citenamefont {Arlt}, \citenamefont {Ertmer},
  \citenamefont {Lisdat}, \citenamefont {Santos}, \citenamefont {Smerzi} \emph
  {et~al.}}]{kruse2016improvement}%
  \BibitemOpen
  \bibfield  {author} {\bibinfo {author} {\bibfnamefont {I.}~\bibnamefont
  {Kruse}}, \bibinfo {author} {\bibfnamefont {K.}~\bibnamefont {Lange}},
  \bibinfo {author} {\bibfnamefont {J.}~\bibnamefont {Peise}}, \bibinfo
  {author} {\bibfnamefont {B.}~\bibnamefont {L{\"u}cke}}, \bibinfo {author}
  {\bibfnamefont {L.}~\bibnamefont {Pezz{\`e}}}, \bibinfo {author}
  {\bibfnamefont {J.}~\bibnamefont {Arlt}}, \bibinfo {author} {\bibfnamefont
  {W.}~\bibnamefont {Ertmer}}, \bibinfo {author} {\bibfnamefont
  {C.}~\bibnamefont {Lisdat}}, \bibinfo {author} {\bibfnamefont
  {L.}~\bibnamefont {Santos}}, \bibinfo {author} {\bibfnamefont
  {A.}~\bibnamefont {Smerzi}}, \emph {et~al.},\ }\href@noop {} {\bibfield
  {journal} {\bibinfo  {journal} {Physical review letters}\ }\textbf {\bibinfo
  {volume} {117}},\ \bibinfo {pages} {143004} (\bibinfo {year}
  {2016})}\BibitemShut {NoStop}%
\bibitem [{\citenamefont {Cox}\ \emph {et~al.}(2016)\citenamefont {Cox},
  \citenamefont {Greve}, \citenamefont {Weiner},\ and\ \citenamefont
  {Thompson}}]{cox2016deterministic}%
  \BibitemOpen
  \bibfield  {author} {\bibinfo {author} {\bibfnamefont {K.~C.}\ \bibnamefont
  {Cox}}, \bibinfo {author} {\bibfnamefont {G.~P.}\ \bibnamefont {Greve}},
  \bibinfo {author} {\bibfnamefont {J.~M.}\ \bibnamefont {Weiner}},\ and\
  \bibinfo {author} {\bibfnamefont {J.~K.}\ \bibnamefont {Thompson}},\
  }\href@noop {} {\bibfield  {journal} {\bibinfo  {journal} {Physical review
  letters}\ }\textbf {\bibinfo {volume} {116}},\ \bibinfo {pages} {093602}
  (\bibinfo {year} {2016})}\BibitemShut {NoStop}%
\bibitem [{\citenamefont {Hosten}\ \emph {et~al.}(2016)\citenamefont {Hosten},
  \citenamefont {Engelsen}, \citenamefont {Krishnakumar},\ and\ \citenamefont
  {Kasevich}}]{hosten2016measurement}%
  \BibitemOpen
  \bibfield  {author} {\bibinfo {author} {\bibfnamefont {O.}~\bibnamefont
  {Hosten}}, \bibinfo {author} {\bibfnamefont {N.~J.}\ \bibnamefont
  {Engelsen}}, \bibinfo {author} {\bibfnamefont {R.}~\bibnamefont
  {Krishnakumar}},\ and\ \bibinfo {author} {\bibfnamefont {M.~A.}\ \bibnamefont
  {Kasevich}},\ }\href@noop {} {\bibfield  {journal} {\bibinfo  {journal}
  {Nature}\ }\textbf {\bibinfo {volume} {529}},\ \bibinfo {pages} {505}
  (\bibinfo {year} {2016})}\BibitemShut {NoStop}%
\bibitem [{\citenamefont {Luo}\ \emph {et~al.}(2017)\citenamefont {Luo},
  \citenamefont {Zou}, \citenamefont {Wu}, \citenamefont {Liu}, \citenamefont
  {Han}, \citenamefont {Tey},\ and\ \citenamefont
  {You}}]{luo2017deterministic}%
  \BibitemOpen
  \bibfield  {author} {\bibinfo {author} {\bibfnamefont {X.-Y.}\ \bibnamefont
  {Luo}}, \bibinfo {author} {\bibfnamefont {Y.-Q.}\ \bibnamefont {Zou}},
  \bibinfo {author} {\bibfnamefont {L.-N.}\ \bibnamefont {Wu}}, \bibinfo
  {author} {\bibfnamefont {Q.}~\bibnamefont {Liu}}, \bibinfo {author}
  {\bibfnamefont {M.-F.}\ \bibnamefont {Han}}, \bibinfo {author} {\bibfnamefont
  {M.~K.}\ \bibnamefont {Tey}},\ and\ \bibinfo {author} {\bibfnamefont
  {L.}~\bibnamefont {You}},\ }\href@noop {} {\bibfield  {journal} {\bibinfo
  {journal} {Science}\ }\textbf {\bibinfo {volume} {355}},\ \bibinfo {pages}
  {620} (\bibinfo {year} {2017})}\BibitemShut {NoStop}%
\bibitem [{\citenamefont {Mason}\ \emph {et~al.}(2019)\citenamefont {Mason},
  \citenamefont {Chen}, \citenamefont {Rossi}, \citenamefont {Tsaturyan},\ and\
  \citenamefont {Schliesser}}]{mason2019continuous}%
  \BibitemOpen
  \bibfield  {author} {\bibinfo {author} {\bibfnamefont {D.}~\bibnamefont
  {Mason}}, \bibinfo {author} {\bibfnamefont {J.}~\bibnamefont {Chen}},
  \bibinfo {author} {\bibfnamefont {M.}~\bibnamefont {Rossi}}, \bibinfo
  {author} {\bibfnamefont {Y.}~\bibnamefont {Tsaturyan}},\ and\ \bibinfo
  {author} {\bibfnamefont {A.}~\bibnamefont {Schliesser}},\ }\href@noop {}
  {\bibfield  {journal} {\bibinfo  {journal} {Nature Physics}\ }\textbf
  {\bibinfo {volume} {15}},\ \bibinfo {pages} {745} (\bibinfo {year}
  {2019})}\BibitemShut {NoStop}%
\bibitem [{\citenamefont {Bao}\ \emph {et~al.}(2020)\citenamefont {Bao},
  \citenamefont {Duan}, \citenamefont {Jin}, \citenamefont {Lu}, \citenamefont
  {Li}, \citenamefont {Qu}, \citenamefont {Wang}, \citenamefont {Novikova},
  \citenamefont {Mikhailov}, \citenamefont {Zhao} \emph
  {et~al.}}]{bao2020spin}%
  \BibitemOpen
  \bibfield  {author} {\bibinfo {author} {\bibfnamefont {H.}~\bibnamefont
  {Bao}}, \bibinfo {author} {\bibfnamefont {J.}~\bibnamefont {Duan}}, \bibinfo
  {author} {\bibfnamefont {S.}~\bibnamefont {Jin}}, \bibinfo {author}
  {\bibfnamefont {X.}~\bibnamefont {Lu}}, \bibinfo {author} {\bibfnamefont
  {P.}~\bibnamefont {Li}}, \bibinfo {author} {\bibfnamefont {W.}~\bibnamefont
  {Qu}}, \bibinfo {author} {\bibfnamefont {M.}~\bibnamefont {Wang}}, \bibinfo
  {author} {\bibfnamefont {I.}~\bibnamefont {Novikova}}, \bibinfo {author}
  {\bibfnamefont {E.~E.}\ \bibnamefont {Mikhailov}}, \bibinfo {author}
  {\bibfnamefont {K.-F.}\ \bibnamefont {Zhao}}, \emph {et~al.},\ }\href@noop {}
  {\bibfield  {journal} {\bibinfo  {journal} {Nature}\ }\textbf {\bibinfo
  {volume} {581}},\ \bibinfo {pages} {159} (\bibinfo {year}
  {2020})}\BibitemShut {NoStop}%
\bibitem [{\citenamefont {Pedrozo-Pe{\~n}afiel}\ \emph
  {et~al.}(2020)\citenamefont {Pedrozo-Pe{\~n}afiel}, \citenamefont {Colombo},
  \citenamefont {Shu}, \citenamefont {Adiyatullin}, \citenamefont {Li},
  \citenamefont {Mendez}, \citenamefont {Braverman}, \citenamefont {Kawasaki},
  \citenamefont {Akamatsu}, \citenamefont {Xiao} \emph
  {et~al.}}]{pedrozo2020entanglement}%
  \BibitemOpen
  \bibfield  {author} {\bibinfo {author} {\bibfnamefont {E.}~\bibnamefont
  {Pedrozo-Pe{\~n}afiel}}, \bibinfo {author} {\bibfnamefont {S.}~\bibnamefont
  {Colombo}}, \bibinfo {author} {\bibfnamefont {C.}~\bibnamefont {Shu}},
  \bibinfo {author} {\bibfnamefont {A.~F.}\ \bibnamefont {Adiyatullin}},
  \bibinfo {author} {\bibfnamefont {Z.}~\bibnamefont {Li}}, \bibinfo {author}
  {\bibfnamefont {E.}~\bibnamefont {Mendez}}, \bibinfo {author} {\bibfnamefont
  {B.}~\bibnamefont {Braverman}}, \bibinfo {author} {\bibfnamefont
  {A.}~\bibnamefont {Kawasaki}}, \bibinfo {author} {\bibfnamefont
  {D.}~\bibnamefont {Akamatsu}}, \bibinfo {author} {\bibfnamefont
  {Y.}~\bibnamefont {Xiao}}, \emph {et~al.},\ }\href@noop {} {\bibfield
  {journal} {\bibinfo  {journal} {Nature}\ }\textbf {\bibinfo {volume} {588}},\
  \bibinfo {pages} {414} (\bibinfo {year} {2020})}\BibitemShut {NoStop}%
\bibitem [{\citenamefont {Shaji}\ and\ \citenamefont
  {Caves}(2007)}]{PhysRevA.76.032111}%
  \BibitemOpen
  \bibfield  {author} {\bibinfo {author} {\bibfnamefont {A.}~\bibnamefont
  {Shaji}}\ and\ \bibinfo {author} {\bibfnamefont {C.~M.}\ \bibnamefont
  {Caves}},\ }\href {https://doi.org/10.1103/PhysRevA.76.032111} {\bibfield
  {journal} {\bibinfo  {journal} {Phys. Rev. A}\ }\textbf {\bibinfo {volume}
  {76}},\ \bibinfo {pages} {032111} (\bibinfo {year} {2007})}\BibitemShut
  {NoStop}%
\bibitem [{\citenamefont {Matsuzaki}\ \emph {et~al.}(2011)\citenamefont
  {Matsuzaki}, \citenamefont {Benjamin},\ and\ \citenamefont
  {Fitzsimons}}]{matsuzaki2011magnetic}%
  \BibitemOpen
  \bibfield  {author} {\bibinfo {author} {\bibfnamefont {Y.}~\bibnamefont
  {Matsuzaki}}, \bibinfo {author} {\bibfnamefont {S.~C.}\ \bibnamefont
  {Benjamin}},\ and\ \bibinfo {author} {\bibfnamefont {J.}~\bibnamefont
  {Fitzsimons}},\ }\href@noop {} {\bibfield  {journal} {\bibinfo  {journal}
  {Physical Review A}\ }\textbf {\bibinfo {volume} {84}},\ \bibinfo {pages}
  {012103} (\bibinfo {year} {2011})}\BibitemShut {NoStop}%
\bibitem [{\citenamefont {Chin}\ \emph {et~al.}(2012)\citenamefont {Chin},
  \citenamefont {Huelga},\ and\ \citenamefont {Plenio}}]{chin2012quantum}%
  \BibitemOpen
  \bibfield  {author} {\bibinfo {author} {\bibfnamefont {A.~W.}\ \bibnamefont
  {Chin}}, \bibinfo {author} {\bibfnamefont {S.~F.}\ \bibnamefont {Huelga}},\
  and\ \bibinfo {author} {\bibfnamefont {M.~B.}\ \bibnamefont {Plenio}},\
  }\href@noop {} {\bibfield  {journal} {\bibinfo  {journal} {Physical review
  letters}\ }\textbf {\bibinfo {volume} {109}},\ \bibinfo {pages} {233601}
  (\bibinfo {year} {2012})}\BibitemShut {NoStop}%
\bibitem [{\citenamefont {Demkowicz-Dobrza{\'n}ski}\ \emph
  {et~al.}(2012)\citenamefont {Demkowicz-Dobrza{\'n}ski}, \citenamefont
  {Ko{\l}ody{\'n}ski},\ and\ \citenamefont
  {Gu{\c{t}}{\u{a}}}}]{demkowicz2012elusive}%
  \BibitemOpen
  \bibfield  {author} {\bibinfo {author} {\bibfnamefont {R.}~\bibnamefont
  {Demkowicz-Dobrza{\'n}ski}}, \bibinfo {author} {\bibfnamefont
  {J.}~\bibnamefont {Ko{\l}ody{\'n}ski}},\ and\ \bibinfo {author}
  {\bibfnamefont {M.}~\bibnamefont {Gu{\c{t}}{\u{a}}}},\ }\href@noop {}
  {\bibfield  {journal} {\bibinfo  {journal} {Nature communications}\ }\textbf
  {\bibinfo {volume} {3}},\ \bibinfo {pages} {1} (\bibinfo {year}
  {2012})}\BibitemShut {NoStop}%
\bibitem [{\citenamefont {Chaves}\ \emph {et~al.}(2013)\citenamefont {Chaves},
  \citenamefont {Brask}, \citenamefont {Markiewicz}, \citenamefont
  {Ko{\l}ody{\'n}ski},\ and\ \citenamefont {Ac{\'\i}n}}]{chaves2013noisy}%
  \BibitemOpen
  \bibfield  {author} {\bibinfo {author} {\bibfnamefont {R.}~\bibnamefont
  {Chaves}}, \bibinfo {author} {\bibfnamefont {J.}~\bibnamefont {Brask}},
  \bibinfo {author} {\bibfnamefont {M.}~\bibnamefont {Markiewicz}}, \bibinfo
  {author} {\bibfnamefont {J.}~\bibnamefont {Ko{\l}ody{\'n}ski}},\ and\
  \bibinfo {author} {\bibfnamefont {A.}~\bibnamefont {Ac{\'\i}n}},\ }\href@noop
  {} {\bibfield  {journal} {\bibinfo  {journal} {Physical review letters}\
  }\textbf {\bibinfo {volume} {111}},\ \bibinfo {pages} {120401} (\bibinfo
  {year} {2013})}\BibitemShut {NoStop}%
\bibitem [{\citenamefont {Tanaka}\ \emph {et~al.}(2015)\citenamefont {Tanaka},
  \citenamefont {Knott}, \citenamefont {Matsuzaki}, \citenamefont {Dooley},
  \citenamefont {Yamaguchi}, \citenamefont {Munro},\ and\ \citenamefont
  {Saito}}]{tanaka2015proposed}%
  \BibitemOpen
  \bibfield  {author} {\bibinfo {author} {\bibfnamefont {T.}~\bibnamefont
  {Tanaka}}, \bibinfo {author} {\bibfnamefont {P.}~\bibnamefont {Knott}},
  \bibinfo {author} {\bibfnamefont {Y.}~\bibnamefont {Matsuzaki}}, \bibinfo
  {author} {\bibfnamefont {S.}~\bibnamefont {Dooley}}, \bibinfo {author}
  {\bibfnamefont {H.}~\bibnamefont {Yamaguchi}}, \bibinfo {author}
  {\bibfnamefont {W.~J.}\ \bibnamefont {Munro}},\ and\ \bibinfo {author}
  {\bibfnamefont {S.}~\bibnamefont {Saito}},\ }\href@noop {} {\bibfield
  {journal} {\bibinfo  {journal} {Physical review letters}\ }\textbf {\bibinfo
  {volume} {115}},\ \bibinfo {pages} {170801} (\bibinfo {year}
  {2015})}\BibitemShut {NoStop}%
\bibitem [{\citenamefont {Davis}\ \emph {et~al.}(2016)\citenamefont {Davis},
  \citenamefont {Bentsen},\ and\ \citenamefont
  {Schleier-Smith}}]{davis2016approaching}%
  \BibitemOpen
  \bibfield  {author} {\bibinfo {author} {\bibfnamefont {E.}~\bibnamefont
  {Davis}}, \bibinfo {author} {\bibfnamefont {G.}~\bibnamefont {Bentsen}},\
  and\ \bibinfo {author} {\bibfnamefont {M.}~\bibnamefont {Schleier-Smith}},\
  }\href@noop {} {\bibfield  {journal} {\bibinfo  {journal} {Physical review
  letters}\ }\textbf {\bibinfo {volume} {116}},\ \bibinfo {pages} {053601}
  (\bibinfo {year} {2016})}\BibitemShut {NoStop}%
\bibitem [{\citenamefont {Matsuzaki}\ \emph
  {et~al.}(2018{\natexlab{a}})\citenamefont {Matsuzaki}, \citenamefont
  {Benjamin}, \citenamefont {Nakayama}, \citenamefont {Saito},\ and\
  \citenamefont {Munro}}]{matsuzaki2018quantum1}%
  \BibitemOpen
  \bibfield  {author} {\bibinfo {author} {\bibfnamefont {Y.}~\bibnamefont
  {Matsuzaki}}, \bibinfo {author} {\bibfnamefont {S.}~\bibnamefont {Benjamin}},
  \bibinfo {author} {\bibfnamefont {S.}~\bibnamefont {Nakayama}}, \bibinfo
  {author} {\bibfnamefont {S.}~\bibnamefont {Saito}},\ and\ \bibinfo {author}
  {\bibfnamefont {W.~J.}\ \bibnamefont {Munro}},\ }\href@noop {} {\bibfield
  {journal} {\bibinfo  {journal} {Physical review letters}\ }\textbf {\bibinfo
  {volume} {120}},\ \bibinfo {pages} {140501} (\bibinfo {year}
  {2018}{\natexlab{a}})}\BibitemShut {NoStop}%
\bibitem [{\citenamefont {Ho}\ \emph {et~al.}(2020)\citenamefont {Ho},
  \citenamefont {Hakoshima}, \citenamefont {Matsuzaki}, \citenamefont
  {Matsuzaki},\ and\ \citenamefont {Kondo}}]{hakoshima2020multiparameter}%
  \BibitemOpen
  \bibfield  {author} {\bibinfo {author} {\bibfnamefont {L.~B.}\ \bibnamefont
  {Ho}}, \bibinfo {author} {\bibfnamefont {H.}~\bibnamefont {Hakoshima}},
  \bibinfo {author} {\bibfnamefont {Y.}~\bibnamefont {Matsuzaki}}, \bibinfo
  {author} {\bibfnamefont {M.}~\bibnamefont {Matsuzaki}},\ and\ \bibinfo
  {author} {\bibfnamefont {Y.}~\bibnamefont {Kondo}},\ }\href@noop {}
  {\bibfield  {journal} {\bibinfo  {journal} {Physical Review A}\ }\textbf
  {\bibinfo {volume} {102}},\ \bibinfo {pages} {022602} (\bibinfo {year}
  {2020})}\BibitemShut {NoStop}%
\bibitem [{\citenamefont {Yoshinaga}\ \emph {et~al.}(2021)\citenamefont
  {Yoshinaga}, \citenamefont {Tatsuta},\ and\ \citenamefont
  {Matsuzaki}}]{yoshinaga2021entanglement}%
  \BibitemOpen
  \bibfield  {author} {\bibinfo {author} {\bibfnamefont {A.}~\bibnamefont
  {Yoshinaga}}, \bibinfo {author} {\bibfnamefont {M.}~\bibnamefont {Tatsuta}},\
  and\ \bibinfo {author} {\bibfnamefont {Y.}~\bibnamefont {Matsuzaki}},\
  }\href@noop {} {\bibfield  {journal} {\bibinfo  {journal} {arXiv preprint
  arXiv:2101.02998}\ } (\bibinfo {year} {2021})}\BibitemShut {NoStop}%
\bibitem [{\citenamefont {Kessler}\ \emph {et~al.}(2014)\citenamefont
  {Kessler}, \citenamefont {Lovchinsky}, \citenamefont {Sushkov},\ and\
  \citenamefont {Lukin}}]{kessler2014quantum}%
  \BibitemOpen
  \bibfield  {author} {\bibinfo {author} {\bibfnamefont {E.~M.}\ \bibnamefont
  {Kessler}}, \bibinfo {author} {\bibfnamefont {I.}~\bibnamefont {Lovchinsky}},
  \bibinfo {author} {\bibfnamefont {A.~O.}\ \bibnamefont {Sushkov}},\ and\
  \bibinfo {author} {\bibfnamefont {M.~D.}\ \bibnamefont {Lukin}},\ }\href@noop
  {} {\bibfield  {journal} {\bibinfo  {journal} {Physical review letters}\
  }\textbf {\bibinfo {volume} {112}},\ \bibinfo {pages} {150802} (\bibinfo
  {year} {2014})}\BibitemShut {NoStop}%
\bibitem [{\citenamefont {Arrad}\ \emph {et~al.}(2014)\citenamefont {Arrad},
  \citenamefont {Vinkler}, \citenamefont {Aharonov},\ and\ \citenamefont
  {Retzker}}]{arrad2014increasing}%
  \BibitemOpen
  \bibfield  {author} {\bibinfo {author} {\bibfnamefont {G.}~\bibnamefont
  {Arrad}}, \bibinfo {author} {\bibfnamefont {Y.}~\bibnamefont {Vinkler}},
  \bibinfo {author} {\bibfnamefont {D.}~\bibnamefont {Aharonov}},\ and\
  \bibinfo {author} {\bibfnamefont {A.}~\bibnamefont {Retzker}},\ }\href@noop
  {} {\bibfield  {journal} {\bibinfo  {journal} {Physical review letters}\
  }\textbf {\bibinfo {volume} {112}},\ \bibinfo {pages} {150801} (\bibinfo
  {year} {2014})}\BibitemShut {NoStop}%
\bibitem [{\citenamefont {D{\"u}r}\ \emph {et~al.}(2014)\citenamefont
  {D{\"u}r}, \citenamefont {Skotiniotis}, \citenamefont {Froewis},\ and\
  \citenamefont {Kraus}}]{dur2014improved}%
  \BibitemOpen
  \bibfield  {author} {\bibinfo {author} {\bibfnamefont {W.}~\bibnamefont
  {D{\"u}r}}, \bibinfo {author} {\bibfnamefont {M.}~\bibnamefont
  {Skotiniotis}}, \bibinfo {author} {\bibfnamefont {F.}~\bibnamefont
  {Froewis}},\ and\ \bibinfo {author} {\bibfnamefont {B.}~\bibnamefont
  {Kraus}},\ }\href@noop {} {\bibfield  {journal} {\bibinfo  {journal}
  {Physical Review Letters}\ }\textbf {\bibinfo {volume} {112}},\ \bibinfo
  {pages} {080801} (\bibinfo {year} {2014})}\BibitemShut {NoStop}%
\bibitem [{\citenamefont {Herrera-Mart{\'\i}}\ \emph
  {et~al.}(2015)\citenamefont {Herrera-Mart{\'\i}}, \citenamefont {Gefen},
  \citenamefont {Aharonov}, \citenamefont {Katz},\ and\ \citenamefont
  {Retzker}}]{herrera2015quantum}%
  \BibitemOpen
  \bibfield  {author} {\bibinfo {author} {\bibfnamefont {D.~A.}\ \bibnamefont
  {Herrera-Mart{\'\i}}}, \bibinfo {author} {\bibfnamefont {T.}~\bibnamefont
  {Gefen}}, \bibinfo {author} {\bibfnamefont {D.}~\bibnamefont {Aharonov}},
  \bibinfo {author} {\bibfnamefont {N.}~\bibnamefont {Katz}},\ and\ \bibinfo
  {author} {\bibfnamefont {A.}~\bibnamefont {Retzker}},\ }\href@noop {}
  {\bibfield  {journal} {\bibinfo  {journal} {Physical review letters}\
  }\textbf {\bibinfo {volume} {115}},\ \bibinfo {pages} {200501} (\bibinfo
  {year} {2015})}\BibitemShut {NoStop}%
\bibitem [{\citenamefont {Matsuzaki}\ and\ \citenamefont
  {Benjamin}(2017)}]{matsuzaki2017magnetic}%
  \BibitemOpen
  \bibfield  {author} {\bibinfo {author} {\bibfnamefont {Y.}~\bibnamefont
  {Matsuzaki}}\ and\ \bibinfo {author} {\bibfnamefont {S.}~\bibnamefont
  {Benjamin}},\ }\href@noop {} {\bibfield  {journal} {\bibinfo  {journal}
  {Physical Review A}\ }\textbf {\bibinfo {volume} {95}},\ \bibinfo {pages}
  {032303} (\bibinfo {year} {2017})}\BibitemShut {NoStop}%
\bibitem [{\citenamefont {Unden}\ \emph {et~al.}(2016)\citenamefont {Unden},
  \citenamefont {Balasubramanian}, \citenamefont {Louzon}, \citenamefont
  {Vinkler}, \citenamefont {Plenio}, \citenamefont {Markham}, \citenamefont
  {Twitchen}, \citenamefont {Stacey}, \citenamefont {Lovchinsky}, \citenamefont
  {Sushkov} \emph {et~al.}}]{unden2016quantum}%
  \BibitemOpen
  \bibfield  {author} {\bibinfo {author} {\bibfnamefont {T.}~\bibnamefont
  {Unden}}, \bibinfo {author} {\bibfnamefont {P.}~\bibnamefont
  {Balasubramanian}}, \bibinfo {author} {\bibfnamefont {D.}~\bibnamefont
  {Louzon}}, \bibinfo {author} {\bibfnamefont {Y.}~\bibnamefont {Vinkler}},
  \bibinfo {author} {\bibfnamefont {M.~B.}\ \bibnamefont {Plenio}}, \bibinfo
  {author} {\bibfnamefont {M.}~\bibnamefont {Markham}}, \bibinfo {author}
  {\bibfnamefont {D.}~\bibnamefont {Twitchen}}, \bibinfo {author}
  {\bibfnamefont {A.}~\bibnamefont {Stacey}}, \bibinfo {author} {\bibfnamefont
  {I.}~\bibnamefont {Lovchinsky}}, \bibinfo {author} {\bibfnamefont {A.~O.}\
  \bibnamefont {Sushkov}}, \emph {et~al.},\ }\href@noop {} {\bibfield
  {journal} {\bibinfo  {journal} {Physical review letters}\ }\textbf {\bibinfo
  {volume} {116}},\ \bibinfo {pages} {230502} (\bibinfo {year}
  {2016})}\BibitemShut {NoStop}%
\bibitem [{\citenamefont {Cohen}\ \emph {et~al.}(2016)\citenamefont {Cohen},
  \citenamefont {Pilnyak}, \citenamefont {Istrati}, \citenamefont {Retzker},\
  and\ \citenamefont {Eisenberg}}]{cohen2016demonstration}%
  \BibitemOpen
  \bibfield  {author} {\bibinfo {author} {\bibfnamefont {L.}~\bibnamefont
  {Cohen}}, \bibinfo {author} {\bibfnamefont {Y.}~\bibnamefont {Pilnyak}},
  \bibinfo {author} {\bibfnamefont {D.}~\bibnamefont {Istrati}}, \bibinfo
  {author} {\bibfnamefont {A.}~\bibnamefont {Retzker}},\ and\ \bibinfo {author}
  {\bibfnamefont {H.}~\bibnamefont {Eisenberg}},\ }\href@noop {} {\bibfield
  {journal} {\bibinfo  {journal} {Physical Review A}\ }\textbf {\bibinfo
  {volume} {94}},\ \bibinfo {pages} {012324} (\bibinfo {year}
  {2016})}\BibitemShut {NoStop}%
\bibitem [{\citenamefont {Matsuzaki}\ \emph {et~al.}(2016)\citenamefont
  {Matsuzaki}, \citenamefont {Shimo-Oka}, \citenamefont {Tanaka}, \citenamefont
  {Tokura}, \citenamefont {Semba},\ and\ \citenamefont
  {Mizuochi}}]{matsuzaki2016hybrid}%
  \BibitemOpen
  \bibfield  {author} {\bibinfo {author} {\bibfnamefont {Y.}~\bibnamefont
  {Matsuzaki}}, \bibinfo {author} {\bibfnamefont {T.}~\bibnamefont
  {Shimo-Oka}}, \bibinfo {author} {\bibfnamefont {H.}~\bibnamefont {Tanaka}},
  \bibinfo {author} {\bibfnamefont {Y.}~\bibnamefont {Tokura}}, \bibinfo
  {author} {\bibfnamefont {K.}~\bibnamefont {Semba}},\ and\ \bibinfo {author}
  {\bibfnamefont {N.}~\bibnamefont {Mizuochi}},\ }\href@noop {} {\bibfield
  {journal} {\bibinfo  {journal} {Physical Review A}\ }\textbf {\bibinfo
  {volume} {94}},\ \bibinfo {pages} {052330} (\bibinfo {year}
  {2016})}\BibitemShut {NoStop}%
\bibitem [{\citenamefont {Averin}\ \emph {et~al.}(2016)\citenamefont {Averin},
  \citenamefont {Xu}, \citenamefont {Zhong}, \citenamefont {Song},
  \citenamefont {Wang},\ and\ \citenamefont {Han}}]{averin2016suppression}%
  \BibitemOpen
  \bibfield  {author} {\bibinfo {author} {\bibfnamefont {D.}~\bibnamefont
  {Averin}}, \bibinfo {author} {\bibfnamefont {K.}~\bibnamefont {Xu}}, \bibinfo
  {author} {\bibfnamefont {Y.-P.}\ \bibnamefont {Zhong}}, \bibinfo {author}
  {\bibfnamefont {C.}~\bibnamefont {Song}}, \bibinfo {author} {\bibfnamefont
  {H.}~\bibnamefont {Wang}},\ and\ \bibinfo {author} {\bibfnamefont
  {S.}~\bibnamefont {Han}},\ }\href@noop {} {\bibfield  {journal} {\bibinfo
  {journal} {Physical review letters}\ }\textbf {\bibinfo {volume} {116}},\
  \bibinfo {pages} {010501} (\bibinfo {year} {2016})}\BibitemShut {NoStop}%
\bibitem [{\citenamefont {Beau}\ and\ \citenamefont {del
  Campo}(2017)}]{beau2017nonlinear}%
  \BibitemOpen
  \bibfield  {author} {\bibinfo {author} {\bibfnamefont {M.}~\bibnamefont
  {Beau}}\ and\ \bibinfo {author} {\bibfnamefont {A.}~\bibnamefont {del
  Campo}},\ }\href@noop {} {\bibfield  {journal} {\bibinfo  {journal} {Physical
  review letters}\ }\textbf {\bibinfo {volume} {119}},\ \bibinfo {pages}
  {010403} (\bibinfo {year} {2017})}\BibitemShut {NoStop}%
\bibitem [{\citenamefont {Matsuzaki}\ \emph
  {et~al.}(2018{\natexlab{b}})\citenamefont {Matsuzaki}, \citenamefont
  {Saito},\ and\ \citenamefont {Munro}}]{matsuzaki2018quantum2}%
  \BibitemOpen
  \bibfield  {author} {\bibinfo {author} {\bibfnamefont {Y.}~\bibnamefont
  {Matsuzaki}}, \bibinfo {author} {\bibfnamefont {S.}~\bibnamefont {Saito}},\
  and\ \bibinfo {author} {\bibfnamefont {W.~J.}\ \bibnamefont {Munro}},\
  }\href@noop {} {\bibinfo {title} {Quantum metrology at the heisenberg limit
  with the presence of independent dephasing}} (\bibinfo {year}
  {2018}{\natexlab{b}}),\ \Eprint {https://arxiv.org/abs/1809.00176}
  {arXiv:1809.00176 [quant-ph]} \BibitemShut {NoStop}%
\bibitem [{\citenamefont {Braun}\ and\ \citenamefont
  {Martin}(2011)}]{braun2011heisenberg}%
  \BibitemOpen
  \bibfield  {author} {\bibinfo {author} {\bibfnamefont {D.}~\bibnamefont
  {Braun}}\ and\ \bibinfo {author} {\bibfnamefont {J.}~\bibnamefont {Martin}},\
  }\href@noop {} {\bibfield  {journal} {\bibinfo  {journal} {Nature
  communications}\ }\textbf {\bibinfo {volume} {2}},\ \bibinfo {pages} {1}
  (\bibinfo {year} {2011})}\BibitemShut {NoStop}%
\bibitem [{\citenamefont {Demkowicz-Dobrza{\'n}ski}\ \emph
  {et~al.}(2017)\citenamefont {Demkowicz-Dobrza{\'n}ski}, \citenamefont
  {Czajkowski},\ and\ \citenamefont {Sekatski}}]{demkowicz2017adaptive}%
  \BibitemOpen
  \bibfield  {author} {\bibinfo {author} {\bibfnamefont {R.}~\bibnamefont
  {Demkowicz-Dobrza{\'n}ski}}, \bibinfo {author} {\bibfnamefont
  {J.}~\bibnamefont {Czajkowski}},\ and\ \bibinfo {author} {\bibfnamefont
  {P.}~\bibnamefont {Sekatski}},\ }\href@noop {} {\bibfield  {journal}
  {\bibinfo  {journal} {Physical Review X}\ }\textbf {\bibinfo {volume} {7}},\
  \bibinfo {pages} {041009} (\bibinfo {year} {2017})}\BibitemShut {NoStop}%
\bibitem [{\citenamefont {Sekatski}\ \emph {et~al.}(2017)\citenamefont
  {Sekatski}, \citenamefont {Skotiniotis}, \citenamefont {Ko{\l}ody{\'n}ski},\
  and\ \citenamefont {D{\"u}r}}]{sekatski2017quantum}%
  \BibitemOpen
  \bibfield  {author} {\bibinfo {author} {\bibfnamefont {P.}~\bibnamefont
  {Sekatski}}, \bibinfo {author} {\bibfnamefont {M.}~\bibnamefont
  {Skotiniotis}}, \bibinfo {author} {\bibfnamefont {J.}~\bibnamefont
  {Ko{\l}ody{\'n}ski}},\ and\ \bibinfo {author} {\bibfnamefont
  {W.}~\bibnamefont {D{\"u}r}},\ }\href@noop {} {\bibfield  {journal} {\bibinfo
   {journal} {Quantum}\ }\textbf {\bibinfo {volume} {1}},\ \bibinfo {pages}
  {27} (\bibinfo {year} {2017})}\BibitemShut {NoStop}%
\bibitem [{\citenamefont {Dooley}\ \emph {et~al.}(2018)\citenamefont {Dooley},
  \citenamefont {Hanks}, \citenamefont {Nakayama}, \citenamefont {Munro},\ and\
  \citenamefont {Nemoto}}]{dooley2018robust}%
  \BibitemOpen
  \bibfield  {author} {\bibinfo {author} {\bibfnamefont {S.}~\bibnamefont
  {Dooley}}, \bibinfo {author} {\bibfnamefont {M.}~\bibnamefont {Hanks}},
  \bibinfo {author} {\bibfnamefont {S.}~\bibnamefont {Nakayama}}, \bibinfo
  {author} {\bibfnamefont {W.~J.}\ \bibnamefont {Munro}},\ and\ \bibinfo
  {author} {\bibfnamefont {K.}~\bibnamefont {Nemoto}},\ }\href@noop {}
  {\bibfield  {journal} {\bibinfo  {journal} {npj Quantum Information}\
  }\textbf {\bibinfo {volume} {4}},\ \bibinfo {pages} {1} (\bibinfo {year}
  {2018})}\BibitemShut {NoStop}%
\bibitem [{\citenamefont {Koczor}\ \emph {et~al.}(2020)\citenamefont {Koczor},
  \citenamefont {Endo}, \citenamefont {Jones}, \citenamefont {Matsuzaki},\ and\
  \citenamefont {Benjamin}}]{koczor2020variational}%
  \BibitemOpen
  \bibfield  {author} {\bibinfo {author} {\bibfnamefont {B.}~\bibnamefont
  {Koczor}}, \bibinfo {author} {\bibfnamefont {S.}~\bibnamefont {Endo}},
  \bibinfo {author} {\bibfnamefont {T.}~\bibnamefont {Jones}}, \bibinfo
  {author} {\bibfnamefont {Y.}~\bibnamefont {Matsuzaki}},\ and\ \bibinfo
  {author} {\bibfnamefont {S.~C.}\ \bibnamefont {Benjamin}},\ }\href@noop {}
  {\bibfield  {journal} {\bibinfo  {journal} {New Journal of Physics}\ }\textbf
  {\bibinfo {volume} {22}},\ \bibinfo {pages} {083038} (\bibinfo {year}
  {2020})}\BibitemShut {NoStop}%
\bibitem [{\citenamefont {Mihailov}(1977)}]{mihailov1977addition}%
  \BibitemOpen
  \bibfield  {author} {\bibinfo {author} {\bibfnamefont {V.}~\bibnamefont
  {Mihailov}},\ }\href@noop {} {\bibfield  {journal} {\bibinfo  {journal}
  {Journal of Physics A: Mathematical and General}\ }\textbf {\bibinfo {volume}
  {10}},\ \bibinfo {pages} {147} (\bibinfo {year} {1977})}\BibitemShut
  {NoStop}%
\bibitem [{\citenamefont {Ping}\ \emph {et~al.}(2002)\citenamefont {Ping},
  \citenamefont {Wang},\ and\ \citenamefont {Chen}}]{ping2002group}%
  \BibitemOpen
  \bibfield  {author} {\bibinfo {author} {\bibfnamefont {J.}~\bibnamefont
  {Ping}}, \bibinfo {author} {\bibfnamefont {F.}~\bibnamefont {Wang}},\ and\
  \bibinfo {author} {\bibfnamefont {J.-q.}\ \bibnamefont {Chen}},\ }\href@noop
  {} {\emph {\bibinfo {title} {Group representation theory for physicists}}}\
  (\bibinfo  {publisher} {World Scientific Publishing Company},\ \bibinfo
  {year} {2002})\BibitemShut {NoStop}%
\bibitem [{\citenamefont {Chase}\ and\ \citenamefont
  {Geremia}(2008)}]{chase2008collective}%
  \BibitemOpen
  \bibfield  {author} {\bibinfo {author} {\bibfnamefont {B.~A.}\ \bibnamefont
  {Chase}}\ and\ \bibinfo {author} {\bibfnamefont {J.}~\bibnamefont
  {Geremia}},\ }\href@noop {} {\bibfield  {journal} {\bibinfo  {journal}
  {Physical Review A}\ }\textbf {\bibinfo {volume} {78}},\ \bibinfo {pages}
  {052101} (\bibinfo {year} {2008})}\BibitemShut {NoStop}%
\bibitem [{\citenamefont {Baragiola}\ \emph {et~al.}(2010)\citenamefont
  {Baragiola}, \citenamefont {Chase},\ and\ \citenamefont
  {Geremia}}]{baragiola2010collective}%
  \BibitemOpen
  \bibfield  {author} {\bibinfo {author} {\bibfnamefont {B.~Q.}\ \bibnamefont
  {Baragiola}}, \bibinfo {author} {\bibfnamefont {B.~A.}\ \bibnamefont
  {Chase}},\ and\ \bibinfo {author} {\bibfnamefont {J.}~\bibnamefont
  {Geremia}},\ }\href@noop {} {\bibfield  {journal} {\bibinfo  {journal}
  {Physical Review A}\ }\textbf {\bibinfo {volume} {81}},\ \bibinfo {pages}
  {032104} (\bibinfo {year} {2010})}\BibitemShut {NoStop}%
\end{thebibliography}%
\bibliographystyle{apsrev4-2}

\end{document}